# Maximum Likelihood Estimation for Single Particle, Passive Microrheology Data with Drift

John W. R. Mellnik

*Curriculum in Bioinformatics and Computational Biology, Department of Mathematics, Department of Biomedical Engineering, University of North Carolina at Chapel Hill, Chapel Hill, NC 27599, USA*

*Path BioAnalytics, Inc. Chapel Hill, NC 27510*

Martin Lysy

*Department of Statistics and Actuarial Science, University of Waterloo, Waterloo, ON N2L 3G1, Canada*

Paula A. Vasquez

*Department of Mathematics, University of South Carolina, Columbia, SC 29208, USA*

Natesh S. Pillai

*Department of Statistics, Harvard University, Cambridge, MA 02138, USA*

David B. Hill

*The Marsico Lung Institute, Department of Physics and Astronomy, University of North Carolina at Chapel Hill, Chapel Hill, NC 27599, USA*

Jeremy Cribb

*Department of Physics and Astronomy, University of North Carolina at Chapel Hill, Chapel Hill, NC 27599, USA*

Scott A. McKinley

*Department of Mathematics, Tulane University, New Orleans, LA 70118, USA*

M. Gregory Forest

*Department of Mathematics, Department of Biomedical Engineering, University of North Carolina at Chapel Hill, Chapel Hill, NC 27599, USA*

**Synopsis**

Volume limitations and low yield thresholds of biological fluids have led to widespread use of passive microparticle rheology. The mean-squared-displacement (MSD) statistics of bead position time series (bead paths) are either applied directly to determine the creep compliance [1] or transformed to determine dynamic storage and loss moduli [2]. A prevalent hurdle arises when there is a non-diffusive experimental drift in the data. Commensurate with the magnitude of drift relative to diffusive mobility, quantified by a Péclet number, the MSD statistics are distorted, and thus the path data must be "corrected" for drift. The standard approach is to estimate and subtract the drift from particle paths, and then calculate MSD statistics. We present an alternative, parametric approach using maximum likelihood estimation that *simultaneously* fits drift and diffusive model parameters from the path data; the MSD statistics (and consequently the compliance and dynamic moduli) then follow directly from the best-fit model. We illustrate and compare both methods on simulated path data over a range of Péclet numbers, where exact answers are known. We choose fractional Brownian motion as the numerical model because it affords tunable, sub-diffusive MSD statistics consistent with typical 30 second long, experimental observations of microbeads in several biological fluids. Finally, we apply and compare both methods on data from human bronchial epithelial cell culture mucus.

## I. INTRODUCTION

The primary application that motivates the methods presented here is the determination of diffusive transport properties and linear viscoelastic properties (dynamic storage and loss moduli or creep compliance) of biological soft matter, and human mucus in particular, based on passive particle tracking microrheology. These viscoelastic inferences arise directly from the mean-squared displacement (MSD) of particle position time series (which we refer to as "paths"). For dynamic moduli, the Mason-Weitz protocol [2, 3, 4] or subsequent corrections [5] are applied to ensemble-averaged transforms of the MSD statistics. The creep compliance follows directly from the time-domain MSD statistics [1]. Video microscopy in combination with passive particle tracking (PPT) has been used to explore the physical properties of a wide range of mucus biogels, including cervicovaginal [6, 7, 8] pulmonary [9, 10] and gastrointestinal [11, 12, 13] mucus.

The issue motivating this paper is that often in PPT experiments, the observed particles exhibit drift: a persistent, inadvertent, driven motion potentially due to the light source [10], movement of the optical stage [14, 15], or some other source [16]. We refer to [17] for analysis of static and dynamic errors in PPT data that improve the accuracy of MSD statistics and thereby inferences of material viscoelasticity. For this paper, we restrict ourselves to persistent linear drift over the duration of each particle position time series. Thus the particle observations are a superposition of drift and diffusion. In active biological fluids such as living cells where endogenous DNA domains are fluoresced and tracked, cell translocation and active cellular processes induce additional non-diffusive motion. In viral trafficking within cells, the virus may hijack directed motion along microtubules. Since drift can significantly alter the MSD of tracked particles, and thereby distort the inference of the dynamic moduli or compliance as well as distort the inferred diffusive mobility, the question naturally arises as to how drift should be accounted for in the analysis of PPT data.

In the case of optical stage drift, each particle in the field of view exhibits the same magnitude and direction of movement. Thus, if enough particles are present, the driven motion may be removed by estimating the ensemble average movement of the particles within the field of view and subtracting this drift from each particle path [18]. Other scenarios pose a more difficult challenge due to the potential for temporal and spatial heterogeneity in the drift component of the motion. In highly heterogeneous biological fluids such as mucus, regions of high elasticity, due to high local mucin concentrations, may cause some particles to appear immobile while neighboring particles undergo net transport due to some local flow. In this or analogous scenarios where drift is non-uniform across observed particle paths, if one were to subtract the ensemble-averaged movement of the particles in the field of view from each particle path, one would be *adding* directed motion to the less mobile particles while *subtracting* directed motion from the more mobile particles. In [19] the authors introduce a "de-trending" method for Brownian motion in heterogeneous fluids that is applied to each particle path individually, where the data is reduced by half through restricting analysis to the movement orthogonal to the estimated drift direction. Here we also analyze individual paths, allowing for independent drift per path, for both Brownian and fractional Brownian motion. The reader is referred to a large body of work by Klafter, Metzler, Barkai and collaborators on fractional Brownian motion as well as other sub- and super-diffusive stochastic processes; see the review article by [20]. In [21] a de-trending method is applied to a larger class of diffusive and sub-diffusive processes with drift in two space dimensions, where the full 2d observations are used to simultaneously estimate the diffusive or sub-diffusive model parameters and the drift. The present article aims to introduce and illustrate the parametric, maximum likelihood estimation approach in one-dimensional PPT data; for a more rigorous treatment, including comparisons of competitive models on the basis of the available data, see [21].

We point out that *the debate over the optimal way to remove drift tacitly assumes that directed motion must be removed prior to analyzing the path data*. Historically, this assumption is natural because of the focus on the scaling of the ensemble particle MSD due to purely diffusive dynamics [22, 23, 24, 25], a statistic that can be extremely sensitive to drift [26]. In this article, we take a different approach and show that *deterministic drift does not need to be removed a priori from particle path data to determine the MSD statistics* ***if*** *one posits and exploits a fully parametric statistical model for the underlying drift-diffusion process*. In particular, we focus on fractional Brownian motion (fBm), a parametric statistical model for sub-diffusive processes that has been shown to accurately describe diffusion in mucus gels [9, 21] and other biological soft matter [27, 28]. For other systems where the particle paths do not exhibit relatively uniform power law MSD behavior, our current methods would not be applicable. We are currently considering extensions of our parametric statistical model to other candidate sub-diffusive stochastic models, e.g., generalized Langevin equations that resolve transient sub-diffusion and convergence at long times to normal diffusion. Other types of path properties, such as jumps between different modes of diffusion within a single tracked path, remain for future development.

Using numerical simulations of drift coupled with sub-diffusive fractional Brownian motion (consistent with data from mucus gels), we show that one can easily and accurately estimate the diffusive or sub-diffusive model parameters by maximum likelihood estimation ($MLE$) -- for a wide range of deterministic drift, and without removing the drift *a priori* as is typically done to estimate the MSD statistics of single or ensemble paths. The advantage of $MLE$ arises because drift is treated as a model parameter to be estimated *jointly* with the diffusive or sub-diffusive parameters rather than sequentially, thereby increasing the precision of all parameter estimates. With all parameters thus estimated jointly, it is then straightforward to use fBm parameter estimates to generate the MSD of the purely diffusive dynamics and thereby infer the dynamic viscoelastic moduli or the creep compliance. We illustrate the procedure for a range of Péclet numbers, which we define through a dimensionless ratio of the drift (advection) component relative to the diffusive mobility, for both normal and sub-diffusive processes.

The benefit of numerical simulation is that the *exact* diffusive parameters and viscoelastic properties are known, so that the error (in MSD and properties derived from it) induced by relative drift (parametrized by Péclet number), for any method of estimating these quantities, can be directly measured and compared to any other method. We are thus able to compare the errors in the inference of diffusive process parameters and drift, in dynamic storage and loss moduli, and in creep compliance, among our proposed parametric maximum likelihood estimation method and the standard approach in the passive microrheology literature (based on a least-squares estimate of the MSD after removal of drift). We also show, for posterity, the dramatic errors in the diffusion parameters and viscoelastic properties if one simply ignores the presence of drift. Finally, for experimental illustration purposes, we apply the $MLE$ and standard drift-subtracted MSD estimate approaches to PPT data from human bronchial epithelial cell culture mucus.

The structure of the article is as follows. First, we discuss drawbacks of MSD-based approaches and cursory drift removal. We then introduce a canonical model (fractional Brownian motion) for tunable particle diffusion or sub-diffusion with drift, and provide details on how to simulate particle paths in accordance with this model. Next, we review MSD-based approaches to the recovery of diffusive parameters and present our $MLE$ method. The methods are then compared using simulated data sets for a practical range of Péclet numbers. Finally we illustrate the methods on data from human lung cell culture mucus.

## II. MEAN SQUARED DISPLACEMENT STATISTICS

Given $(M+1)$ observations $X(0), X(\Delta t), X(2\Delta t), \dots, X(M\Delta t)$ of a particle's position, the MSD statistic is calculated as

$$\langle r_{\tau_i} \rangle^2 = \frac{1}{M-i+1} \sum_{j=0}^{M-i} [X(\tau_i + j\Delta t) - X(j\Delta t)]^2, \tag{1}$$

where $\tau_i = i\Delta t$ is called the lag time and $\Delta t$ is the time between observations. Note that for larger lag times, the number of observations decreases, limiting the statistical significance of the MSD. For many diffusive processes, theory and observation suggest that the MSD of particles undergoing diffusion exhibits a power law scaling over a range of lag times for which there are sufficient observations [29, 6, 4, 30, 31, 9]:

$$E[\langle r_\tau \rangle^2] = 2dD\tau^\alpha, \tag{2}$$

where the prefactor $D$ is the "diffusivity" by analogy with simple diffusion, $\alpha$ is a real number in the interval $[0, 2]$, $d$ is the dimensionality, and $E[\dots]$ is notation for the "expected value". For standard Brownian motion without drift, the power-law exponent is $\alpha = 1$. From an accurate estimation of $D$ one infers the fluid viscosity $\eta$ from the Stokes-Einstein relation. [26] illustrated via simulated Brownian motion that linear (i.e. constant) drift causes the log-log plot of MSD versus lag time $\tau$ to tend toward a slope of 2 at large lag times (Figure 1). That is, as $\tau$ increases, $\alpha \rightarrow 2$ and the larger the drift velocity, the smaller the lag time at which this transition occurs.

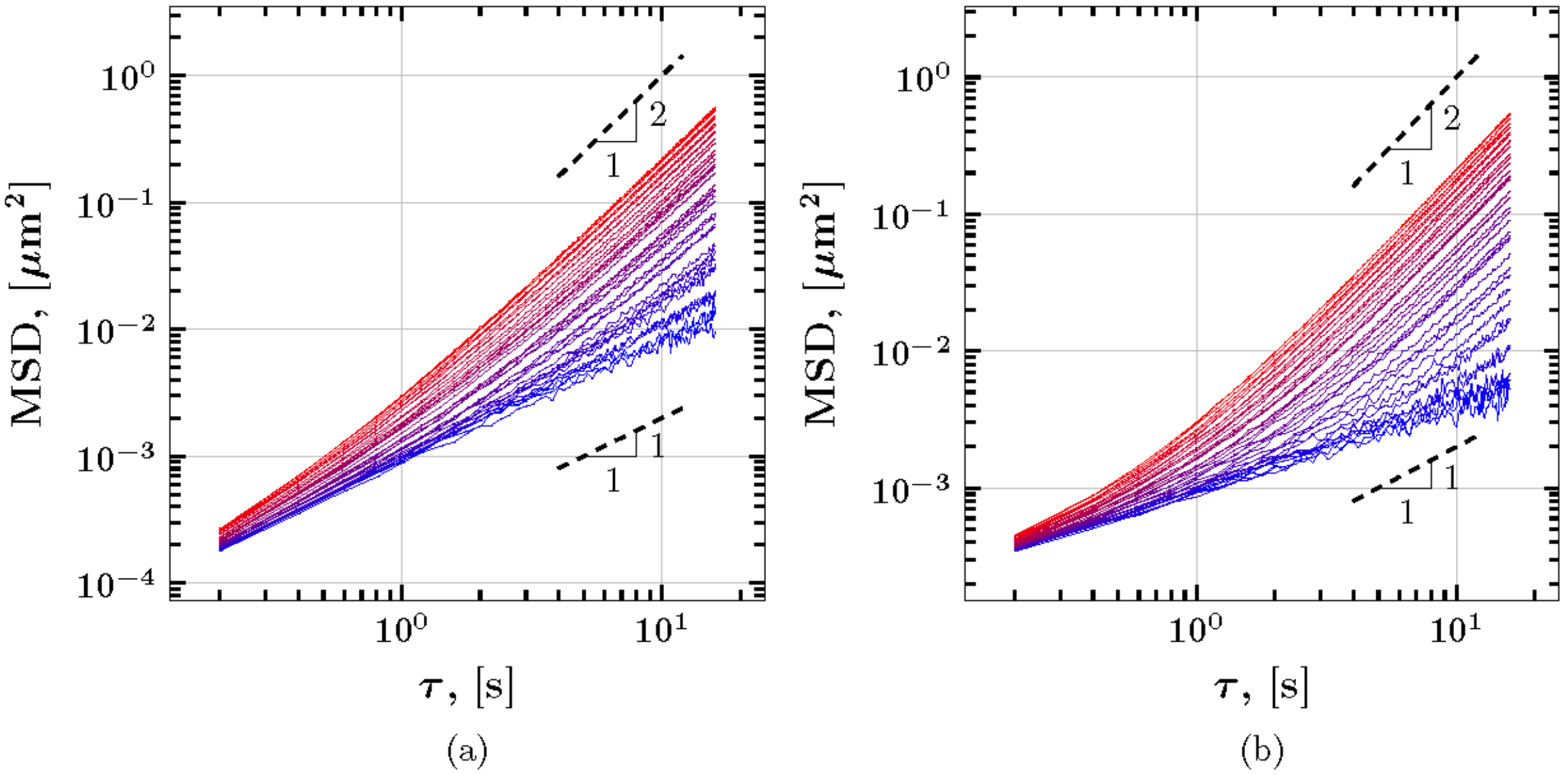

*Figure 1: Path-wise MSD for simulated particles from the Brownian motion (Bm) (a) and fractional Brownian motion (fBm) (b). MSD curves for 40 representative data sets are shown for varying drift characterized by the Péclet number (Pe), the ratio of the advective and diffusive transport rates (Eqn. (13)). The color scheme represents increasing Péclet from* $Pe_{min} = 0$ *(blue) to* $Pe_{max} = 0.73$ *(red), for the Bm data set, and from* $Pe_{min} = 0$ *(blue) to* $Pe_{max} = 0.52$ *(red) for the fBm data set. The upper and lower black dashed lines indicate slopes of 2 (ballistic motion) and 1 (normal diffusion), respectively. The fBm paths (b) are simulated with* $\alpha = 0.6$*; the "diffusivity" pre-factor is chosen to have the same numerical value in the two data sets.*

When one attempts pathwise correction for directed motion by subtracting the mean increment from each path, i.e., the mean of the step-size distribution at the shortest lag time (Figure 2), one inadvertently changes the structure of the entire particle path. That is, *every such modified path is constrained to begin and end at the same spatial position*. To see this, consider a one-dimensional Brownian path, where $X_i = X(i\Delta t)$ is the location of the particle at time $i\Delta t$ with $i = 1, 2, 3 \dots M$. The increments of this process are given by

$$x_i = X_{i+1} - X_i. \tag{3}$$

For Brownian motion, the $x_i$ are normally distributed with mean $\mu\Delta t$ and variance $2D\Delta t$, where $\mu$ is the drift velocity. When no drift is present, the sample mean of $x_i$,

$$\bar{x} = \frac{1}{M-1}\sum_{i=1}^{M-1} x_i, \tag{4}$$

converges to zero as the number of particle positions increases, i.e. as $M \to \infty$. The fact that the distribution of $x_i$ is symmetric with $\bar{x}$ converging to zero intuitively indicates that the particle is expected to make an equal number of steps to the left and right. This however, is *not to say* that a particle diffusing via Brownian motion never travels a net distance. The mean incremental displacement is $\bar{x}$, and when we subtract $\bar{x}$ from each increment, $x_i$, we "snap" the distribution of $X_M$ to zero, inadvertently stipulating that the first and final positions of the particle are the same. Indeed, suppose that $\bar{x}$ is subtracted from each increment to "remove drift," centering the distribution of increments at zero. The resulting modified position process is computed by taking the cumulative sum of the shifted increments, denoted by $\tilde{X}_i$,

$$\begin{aligned}
&\tilde{X}_1 = X_1 \\
&\tilde{X}_2 = X_1 + (x_1 - \bar{x}) \\
&\tilde{X}_3 = X_1 + (x_1 - \bar{x}) + (x_2 - \bar{x}) \\
&\tilde{X}_4 = X_1 + (x_1 - \bar{x}) + (x_2 - \bar{x}) + (x_3 - \bar{x}) \\
&\vdots
\end{aligned} \tag{5}$$

Following this pattern, we collect terms and write the final position $\tilde{X}_M$ as,

$$\tilde{X}_M = X_1 - (M-1)\bar{x} + \sum_{i=1}^{M-1} x_i, \tag{6}$$

which can be simplified further,

$$\tilde{X}_M = X_1 - (M-1)\frac{1}{M-1}\sum_{i=1}^{M-1} x_i + \sum_{i=1}^{M-1} x_i \tag{7}$$

$$\tilde{X}_M = X_1 - \sum_{i=1}^{M-1} x_i + \sum_{i=1}^{M-1} x_i \tag{8}$$

$$\tilde{X}_M = X_1 \tag{9}$$

and thus we see that the *final position has been constrained to the initial position* (Figure 3). It is worth noting that a standard Brownian motion constrained to have predetermined initial and final positions is called a *Brownian bridge* [32], which has completely different correlation structure than the unconstrained motion. While this does not make a difference in the estimation procedure if the paths are from a standard Brownian motion (because the increments are independent), it becomes relevant when the paths are from any stochastic process where the increments may be highly correlated, and in particular, sub-diffusive processes typical of biogels.

An additional drawback of an MSD-based approach to diffusive parameter estimation is the unreliability in the MSD at large lag times. As the lag time increases, the number of increments included in the mean of the squared increments decreases and thus becomes less stable. [26] estimate that only the initial two-thirds of the MSD is statistically reliable. Due to experimental factors (particles exiting the focal plane of the microscope) limiting the ability to collect data over long time scales, the uncertainty in the MSD for large lag times can have a pronounced impact on the accurate recovery of diffusive and viscoelastic properties.

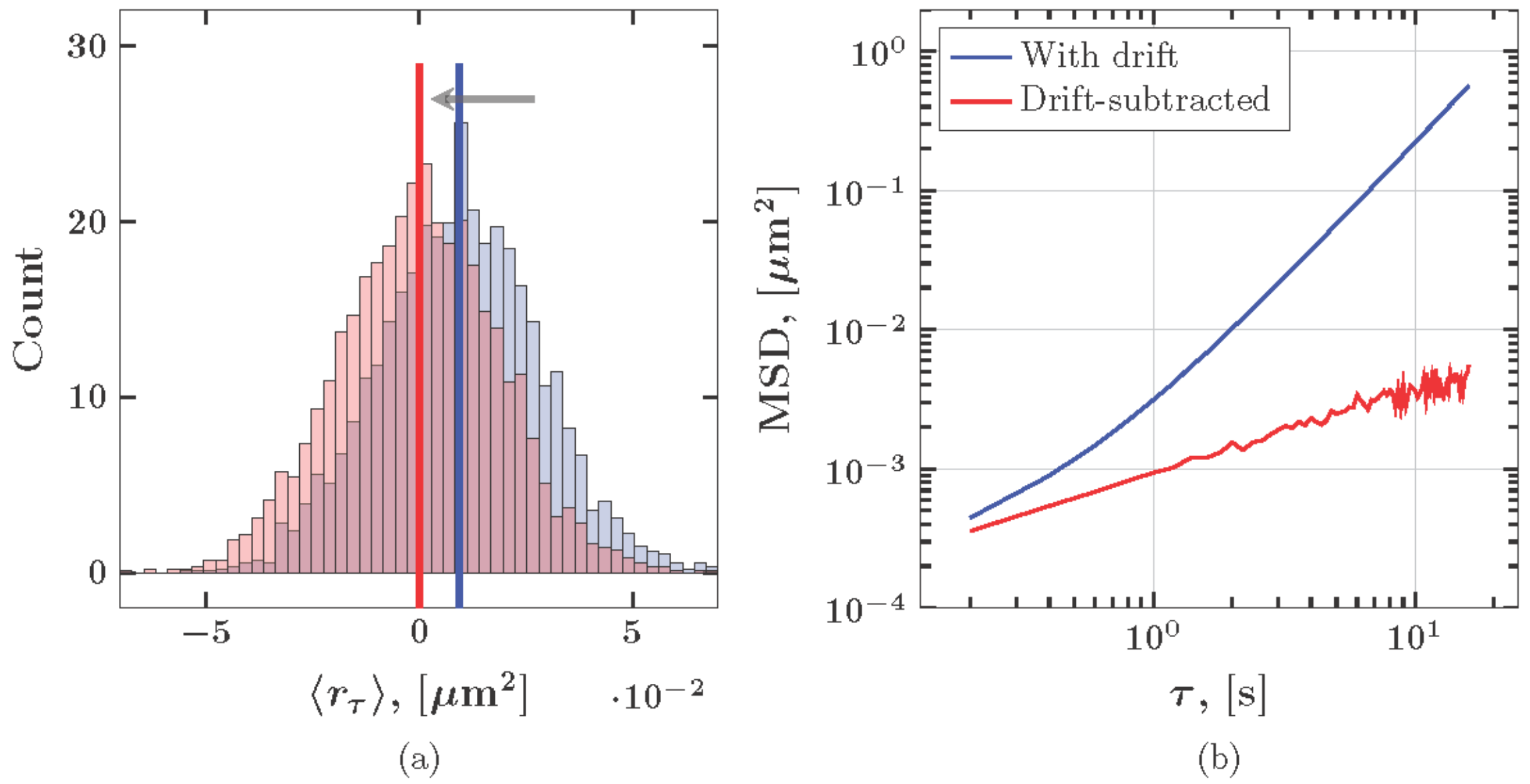

*Figure 2: Impact of drift-subtraction on the distribution of increments and MSD for a representative sub-diffusive fractional Brownian motion path with true parameter values $\alpha = 0.60$ and $D = 4.67 \times 10^{-4}$ $\mu m^2 s^{-\alpha}$. The estimated parameter values based on a simple least-squares fit to the drift subtracted MSD are $\alpha = 0.53$ and $D = 5.30 \times 10^{-4}$ $\mu m^2 s^{-\alpha}$. The distribution of increments (a) is shown at $\tau = 5$ s for a single particle path with $Pe = 0.5$ before (blue) and after (red) drift subtraction. Before drift subtraction, the mean of the distribution of increments (solid blue line) is $9.40 \times 10^{-3}$ $\mu m$. Subtracting drift centers the distribution at zero (solid red line). The MSD is also shown for this path before and after drift subtraction (b).*

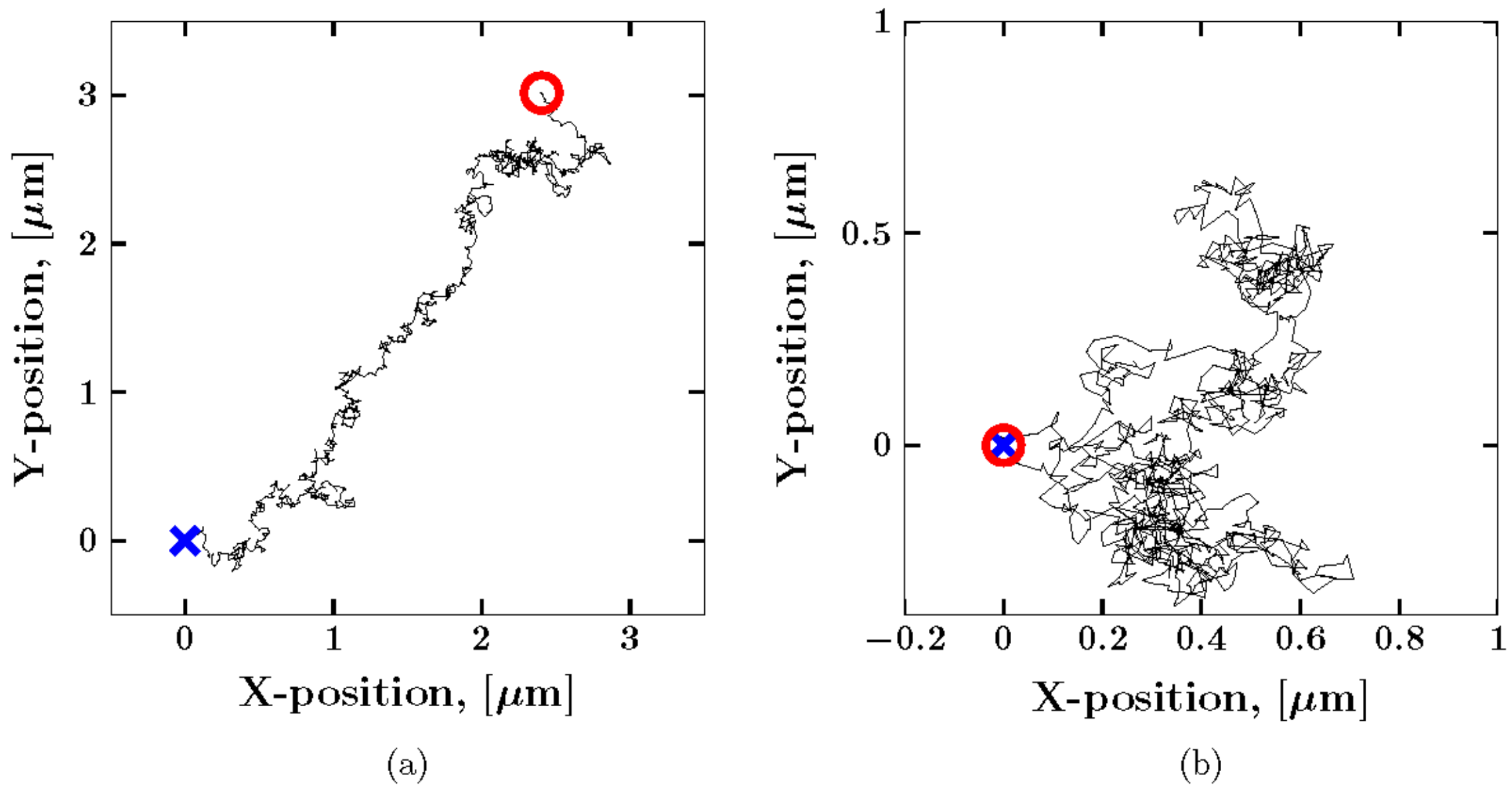

*Figure 3: Sample Brownian path with drift (a) and after the drift has been removed by subtracting the mean displacement from each increment (b). The beginning and end of the path have been marked with a blue x and a red circle, respectively.*

## III. FRACTIONAL BROWNIAN MOTION AND DRIFT

Recently, [21] considered fractional Brownian motion (fBm) as a model for the movement of micron-scale particles over a 30 second observation time at 60 frames per second temporal resolution in human bronchial epithelial mucus. Under this model, the particle's position process $X(t)$ in one dimension is written as the sum of a deterministic term representing the drift and a stochastic term representing the particle's thermally activated diffusive movements:

$$X(t) = \mu t + \sqrt{2D} W_\alpha(t), \tag{10}$$

where $W_\alpha(t)$ is a continuous Gaussian process with mean zero and covariance

$$cov[W_\alpha(t), W_\alpha(s)] = \frac{1}{2}(|t|^\alpha + |s|^\alpha - |t-s|^\alpha), \quad 0 < \alpha < 2. \tag{11}$$

For $\alpha = 1$, Eqn. (10) reduces to Brownian motion with drift, and the increment process is uncorrelated: $cov[x_i, x_{i+k}] = D\Delta t$. For $\alpha \neq 1$, the increment process for fBm has correlation

$$cov[x_i, x_{i+k}] = D\Delta t^{\alpha}(|k+1|^{\alpha} + |k-1|^{\alpha} - 2|k|^{\alpha}). \quad (12)$$

For fBm processes, the MSD has the same scaling relation as Brownian motion [33], i.e., $\langle r_\tau \rangle^2 \sim D\tau^{\alpha}$, although, unlike Brownian motion, the power-law exponent is not necessarily unity. When $\alpha < 1$, the fBm increments are negatively correlated and the position process exhibits subdiffusive behavior. When $\alpha > 1$, the fBm increments are positively correlated and the position process exhibits superdiffusive behavior.

Calculating the increments $x_i$ provides a simple way to estimate the drift exhibited by a particle since the mean, or *expected value* $E[\dots]$, of the increments is $E[x_i] = \mu\Delta t$, for both Brownian and fractional Brownian processes. To generalize our analysis, we characterize results in terms of a Péclet number ($Pe$), a dimensionless ratio of the advective and diffusive transport rates. Given the increments of a particle path computed for a given lag time $\Delta t$, we introduce an approximate Péclet number that is derived directly from the data collected in drift-diffusion microrheology experiments where one does not *a priori* know either the fluid velocity or the fluid diffusivity, and furthermore where fractional diffusivities arise. Namely, at the observation timescale ($\Delta t = 1/60$ s), from the observed data, we define an approximate Péclet number as the ratio of the mean increment of the particle path, $E[x_i]$ or the expected value of $x_i$, and the standard deviation $SD[x_i]$ of the increments,

$$Pe \equiv \frac{E[x_i]}{SD[x_i]}. \quad (13)$$

This definition is not a "pure" Péclet number for all particle tracking experiments. The numerator includes a lower order contribution from apparent diffusive drift due to a finite number of observations. The denominator will include the standard deviation in the advection process if the drift per increment is not identical (due to non-constant drift or measurement error). Nonetheless, this is a reasonable Péclet number based purely on the data, which one could correct later from the results of the MLE method. (GF thanks Ian Seim for a discussion of this issue.)

**IV.A METHODS: Simulation**

To generate a particle path exhibiting linear drift and fractional or normal Brownian dynamics, we first generate the increment process for an fBm path without drift, and then add the desired drift to the path. To generate fBm observations $X_1, \dots, X_M$ for a particular choice of $\alpha$, we first construct the covariance matrix $\boldsymbol{S}$ of the increment process according to Eqn. (12). That is, the $i, j^{th}$ elements of $\boldsymbol{S}$ is

$$\boldsymbol{S}_{i,j} = cov[x_i, x_j] \tag{14}$$

for $i, j = 1, 2, \dots M$. Let $\boldsymbol{LL}' = \boldsymbol{S}$ be the Cholesky decomposition of $\boldsymbol{S}$ and let $\boldsymbol{u}$ be a vector of $M$ independent and identically distributed draws from a standard normal distribution. A simulated particle path is generated as $\boldsymbol{x} = (x_1, \dots, x_M)' = \sqrt{2D}\boldsymbol{Lu}$, $X_j = \sum_{i=1}^{j} x_i$. Using this method, two sets of simulated data are generated. The first set is subdiffusive with $\alpha = 0.6$ and $D = 4.67 \times 10^{-4}\ \mathrm{\mu m^2 s^{-\alpha}}$, mimicking the estimated parameter values based on experimental observations of 1 $\mathrm{\mu m}$ diameter particles in 4 weight percent human bronchial epithelial mucus [9]. The second data set exhibits standard Brownian motion, i.e., $\alpha = 1$, with diffusivity $D = 4.67 \times 10^{-4}\ \mathrm{\mu m^2 s^{-1}}$ corresponding to a $1\ \mathrm{\mu m}$ diameter particle in a fluid with viscosity of $1.86$ Pa s at 23 °C. Each simulated path is generated with a temporal resolution of 5 frames per second and a length of $M = 2{,}992$ steps, mimicking experimental conditions for the experimental data presented in Section VII.

Linear drift is added to the simulated paths by calculating the increments, adding directed motion, then taking the cumulative sum of the result. We simulate both fractional and normal diffusive paths for Péclet numbers in the range $[0, 0.73]$ for Brownian motion and $[0, 0.52]$ for fractional Brownian motion, chosen such that we match the range of drift relative to diffusion as defined by Eqn. (13) for the observed experimental data (Section VII). A position process with linear drift is given by,

$$X_j = \sum_{i=1}^{j} (x_i + \Lambda), \tag{15}$$

where $\Lambda$ is a scaling factor with units of $\mathrm{\mu m}$. We generate 100 simulated paths with drift for $\Lambda$ spanning the interval $[0, 9.34 \times 10^{-3}]$ in increments of $2.3 \times 10^{-4}\ \mathrm{\mu m}$, resulting in 4,100 simulated fBm paths ($\alpha = 0.6$) and 4,100 simulated Brownian paths ($\alpha = 1$). These data sets will be referred to as the fractional Brownian motion (fBm) and Brownian motion (Bm) data sets, respectively.

### IV.B METHODS: Experimental

Mucus harvested from human bronchial epithelia (HBE) cell cultures has proven to be a useful model system for the study of the role of mucus in pulmonary physiology. Briefly, cells are obtained from excess surgical tissue that is procured by the UNC Tissue Core Facility. Cells are then seeded on 0.4 µm Millicell (Millipore, Billerica, MA) coated with collagen. Cells are grown using air liquid interface media as previously described [34]. After 3 weeks of growth, cells are confluent and are ready for mucus harvest as previously described [35, 36, 37, 9]. Biochemically, the composition of mucus harvested from HBE cultures is highly conversed with human sputum [38] and reproduces the osmotic pressure [39] and rheological properties of sputum [9]. Physiologically, HBE mucus has been used to ascertain the role of mucus concentration in pathological bacterial biofilm formation [36], reduced neutrophil activity and motility [35], and clearance [40]. Finally, the concentration of HBE mucus, a simple biochemical property of mucus, has been shown to be correlated with disease states and severity [36, 39, 9, 40].

We selected 1 µm diameter polystyrene particles with carboxyl surface chemistry (Fluospheres, Fisher Scientific) for use in our assays. This particle size is substantially larger than the length scales of the mucin mesh network [41, 42, 43, 44, 9]. Further, the carboxyl functionalization rather than an amine surface chemistry was chosen as previous studies have shown that amine treated beads have impaired diffusion in sputum [45]. PEG surface chemistries, which enhance the diffusion of smaller particles (200 nm and smaller) [41, 46] in mucus have little effect on the diffusivity of larger (>500 nm diameter) particles [46]. These factors lead us to use 1 µm diameter particles with carboxylic acid functionalization. Particles of this size will more faithfully mimic linear macroscopic rheology [43, 47]. Further, by using HBE mucus prepared to concentrations prescribed by physiologically relevant states, we are able to avoid the 100 fold variations in reported literature values that were evident in earlier mucus macroscopic assays [48].

## V. APPROACHES TO PARAMETER ESTIMATION

We consider three approaches to diffusive parameter estimation.

**Simple least squares** Noting that the subdiffusive MSD is linear on the log-log scale,

$$\ln\left(E[\langle r_\tau \rangle_i{}^2]\right) = \ln(2D) + \alpha \ln(\tau_i), \tag{16}$$

a longstanding approach to estimate $D$ and $\alpha$ is to minimize the least squares ($LS$) objective function

$$\sum_{i=1}^{M} (y_i - c - \alpha t_i)^2, \tag{17}$$

in terms of $c$ and $\alpha$ where,

$$y_i = \ln\left[\langle r_\tau \rangle_i{}^2\right], \qquad \mathrm{c} = \ln[2D], \qquad t_i = \ln[\tau_i]. \tag{18}$$

Recall that $M$ is the number of particle positions. The minimum of Eqn. (17) is obtained at $\tilde{\alpha} = \sum_{i=1}^{M} y_i t_i / \sum_{i=1}^{M} t_i^2$ and $\tilde{c} = \bar{y} - \tilde{\alpha}\bar{t}$.

**Drift-Subtracted Least Squares** The drift-subtracted least squares ($DLS$) approach subtracts $\bar{x}$, the mean increment (Eqn. (4)), from each $x_i$ (Eqn. (3)), centering the distribution of increments at zero, dictating the equivalence of the initial and final position, before applying the approach described above for least squares estimation.

**Full Model MLE** This approach applies maximum likelihood estimation ($MLE$) to Eqn. (10) to estimate $\mu$, $D$ and $\alpha$ directly from the raw data without first estimating the MSD statistic. The fBm model in Eqn. (10) specifies that the increments $\boldsymbol{x} = (x_1, \dots, x_M)'$ have a multivariate Gaussian distribution with mean $E[x_i] = \mu\Delta t$ and variance matrix $\boldsymbol{S}$ given by Eqn. (14), denoted

$$\boldsymbol{x} \sim \mathcal{N}(\mu\Delta t, \boldsymbol{S}). \tag{19}$$

Let $\boldsymbol{S} = \sigma^2 \boldsymbol{V}_\alpha$, where $\sigma = 2D$ and $\boldsymbol{V}_\alpha$ is a matrix independent of $D$. The likelihood of a set of model parameters $\boldsymbol{\theta} = (\mu, \sigma, \alpha)$, given a set of observations $\boldsymbol{x}$, is given by the likelihood function

$$\mathcal{L}(\boldsymbol{\theta}|\boldsymbol{x}) = \exp\left[-\frac{1}{2}\frac{(\boldsymbol{x} - \mu\Delta t)'\boldsymbol{V}_\alpha^{-1}(\boldsymbol{x} - \mu\Delta t)}{\sigma^2} - \frac{1}{2}\ln(|\sigma^2 \boldsymbol{V}_\alpha|)\right], \tag{20}$$

which, up to a factor of $(2\pi)^{-M/2}$, is the probability density function (PDF) of the multivariate Gaussian specified by Eqn. (19). The $MLE$ of the parameters is

$$\widehat{\boldsymbol{\theta}} = \mathrm{argmax}_{\boldsymbol{\theta}} \mathcal{L}(\boldsymbol{\theta}|\boldsymbol{x}), \tag{21}$$

the value of $\boldsymbol{\theta}$ that maximizes $\mathcal{L}(\boldsymbol{\theta}|\boldsymbol{x})$. The three-dimensional optimization problem in Eqn. (21) can be reduced to a one-dimensional problem by maximizing in $(\mu, \sigma)$ for fixed $\alpha$. That is, let

$$\boldsymbol{y} = \boldsymbol{y}_\alpha = [\boldsymbol{V}_\alpha]^{-1/2}\boldsymbol{x}, \qquad \text{and} \qquad \boldsymbol{z} = \boldsymbol{z}_\alpha = \Delta t[\boldsymbol{V}_\alpha]^{-1/2}\boldsymbol{1}_M, \tag{22}$$

where $\boldsymbol{1}_M = (1, 1, \dots 1)'$. Then the $\boldsymbol{y} = (y_1, \dots y_M)'$ are independent Gaussians with common variance $\sigma^2$, so that

$$y_i \overset{ind}{\sim} \mathcal{N}(\mu z_i, \sigma^2), \tag{23}$$

such that for fixed $\alpha$, the two-parameter likelihood function $\mathcal{L}_\alpha(\mu, \sigma|\boldsymbol{x})$ is

$$\mathcal{L}_\alpha(\mu, \sigma|\boldsymbol{x}) = \exp\left[-\mathrm{M}\ln(\sigma) - \sum_{i=1}^{M} \frac{(y_i - \mu z_i)^2}{2\sigma^2}\right]. \tag{24}$$

The values $(\hat{\mu}_\alpha, \hat{\sigma}_\alpha)$ that maximize $\mathcal{L}_\alpha(\mu, \sigma|\boldsymbol{x})$ are

$$\hat{\mu}_\alpha = \frac{\sum_{i=1}^M z_i y_i}{\sum_{i=1}^M z_i^2}, \qquad \hat{\sigma}_\alpha = \left( \frac{\sum_{i=1}^M (y_i - \hat{\mu}_\alpha z_i)^2}{M} \right)^{1/2}. \tag{25}$$

The $MLE$ of $\alpha$ for Eqn. (10) is thus obtained by maximizing the one-dimensional *profile likelihood function* [49]

$$\mathcal{L}_{\text{prof}}(\alpha|\boldsymbol{x}) \stackrel{\text{def}}{=} \mathcal{L}(\hat{\mu}_\alpha, \hat{\sigma}_\alpha, \alpha|\boldsymbol{x}). \tag{26}$$

Specifically, by substituting Eqn. (25) into Eqn. (24), we find the $\hat{\alpha}$ that maximizes

$$\ell_{\text{prof}}(\alpha|\boldsymbol{x}) = \ln\left(\mathcal{L}_{\text{prof}}(\alpha|\boldsymbol{x})\right) + C \tag{27}$$

$$= -\frac{1}{2}[M\ln(\hat{\sigma}_\alpha^2) + \ln(|\boldsymbol{V}_\alpha|)]. \tag{28}$$

The resulting parameter estimates $\hat{\theta} = (\hat{\mu}_{\hat{\alpha}}, \hat{\alpha}, \hat{\sigma}_{\hat{\alpha}})$ are precisely those that maximize the full likelihood $\mathcal{L}(\theta|x)$, thereby reducing the numerical optimization problem from three parameters to one. Moreover, we note that for arbitrary variance matrix $\boldsymbol{V}$, the linear systems in Eqn. (22) are solved in $O(M^3)$ operations. However, since $\boldsymbol{V}_\alpha$ is a Toeplitz matrix [50], the systems can be solved in $O(M^2)$ operations using the Durbin-Levinson algorithm [51, 52].

The $MLE$ was implemented in MATLAB, except the Durbin-Levinson algorithm which was implemented in C++. While the DLS estimate is much faster to compute, both algorithms scale as $O(M^2)$. For 2-d paths of length $M = 3000, 5000, 10000$, each $MLE$ took $0.3, 0.8, 3.3$ seconds to evaluate on a personal computer. Pseudocode for the $MLE$ of fBm with drift can be found in Appendix A.

Much like the least squares approach involving the sample MSD, the maximum likelihood approach we have described hinges on the minimization of a quadratic objective function. However, whereas the least squares approach estimates the drift only once, the $MLE$ estimates the "optimal" drift and diffusivity for every value of $\alpha$. That is, the least squares estimate of the drift by $\bar{x}$ would be optimal if the increments were uncorrelated, whereas the $MLE$ approach estimates the drift by a weighted average of the increments, $\hat{\mu}_\alpha$ that accounts for their correlation. Indeed, $\hat{\mu}_\alpha = \bar{x}$ only when fBm reduces to ordinary diffusion with $\alpha = 1$.

### VI.A. RESULTS: Simulated Data

For each simulated path, we compute the path-wise MSD given by Eqn. (1). To estimate the viscous and elastic moduli, we follow [2] where the complex modulus is

$$G^*(\omega) = i\omega\eta^*(\omega) = G'(\omega) + iG''(\omega) = \frac{k_B T}{\pi a i \omega \mathfrak{F}\{\langle r_\tau \rangle^2\}}, \tag{29}$$

where $\mathfrak{F}\{g(\tau)\} = \int g(\tau)e^{-\omega i \cdot \tau}d\tau$ denotes the Fourier transform. Note, the dynamic viscosity $\eta'(\omega)$ is related to the viscous modulus via $\omega\eta'(\omega) = G''(\omega)$. We note further that for fBM and our experimental data, the MSD curves are highly uniform. For data where the MSD curves are non-uniform, corrections to (20) should be used, *cf.* [5].

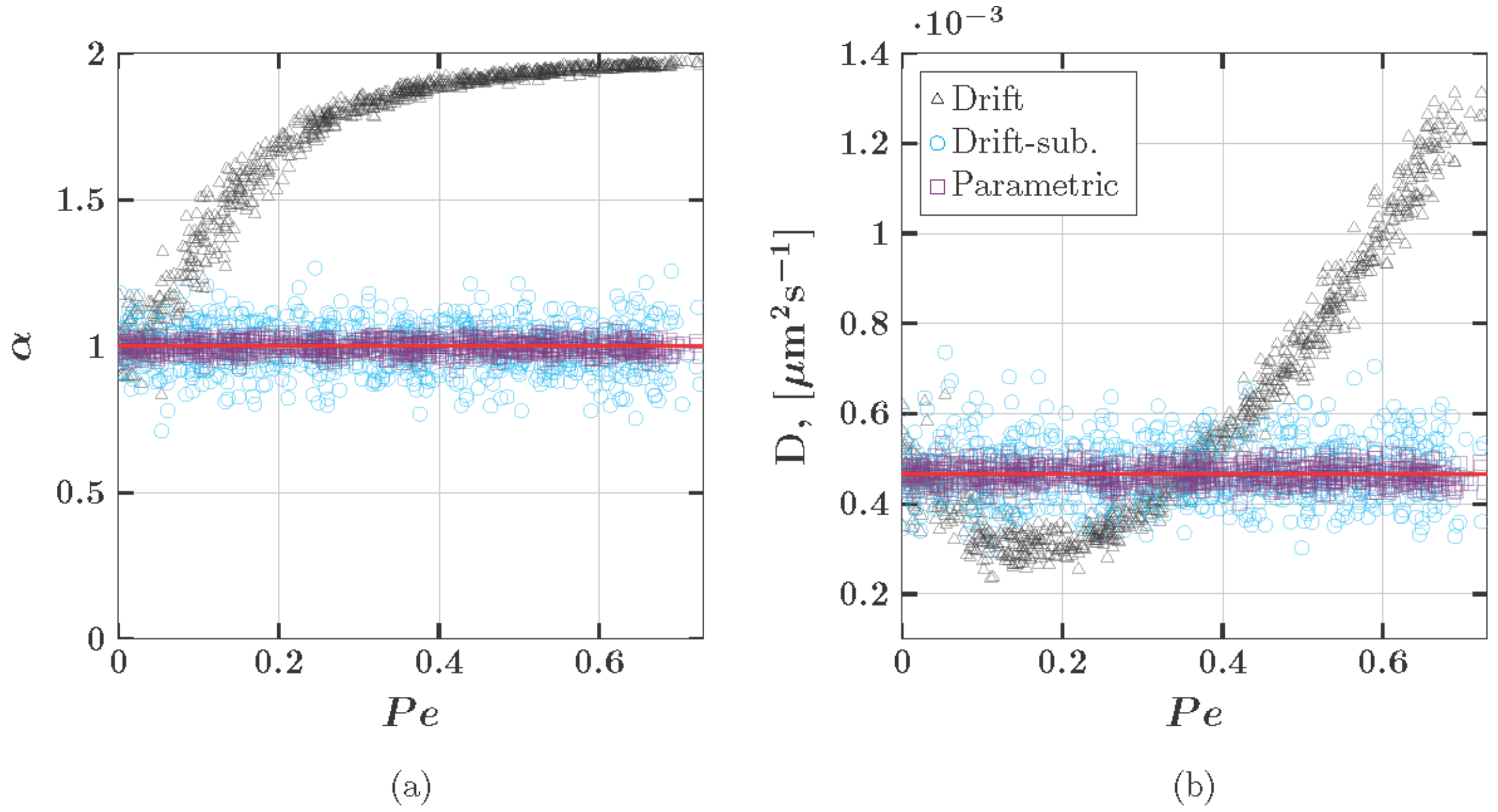

*Figure 4: Estimated values of $\alpha$ (a) and $D$ (b) ignoring drift (black circles), subtracting drift (blue circles) and our maximum likelihood method (violet circles). The true values of $\alpha$ and $D$ are shown in solid red lines. The true values of each parameter are $\alpha = 1$ and $D = 4.67 \times 10^{-4}\ \mu m^2 s^{-1}$.*

Figure 4 shows pathwise estimates of the diffusivity $D$ and the power-law exponent $\alpha$ as a function of the Péclet number ($Pe$) over the range $Pe = [0, 0.73]$ using the three methods described in Section V. The solid red line in each panel is the value used to generate the simulated data, which is reasonably recovered by each technique when no drift is present, i.e. $Pe = 0$. For Brownian paths when drift is present, ignoring drift completely leads to dramatically incorrect results. In Figure 5, the relative error in the estimation of the viscosity for the Bm data found by applying the Stokes-Einstein relation is reported for each estimation approach. The mean relative error in the estimation of the viscosity when not accounting for drift is 39.5%. When applying a drift-subtracted least squares ($DLS$) approach and a parametric $MLE$ approach the mean relative error is 11.1% and 3.6%, respectively.

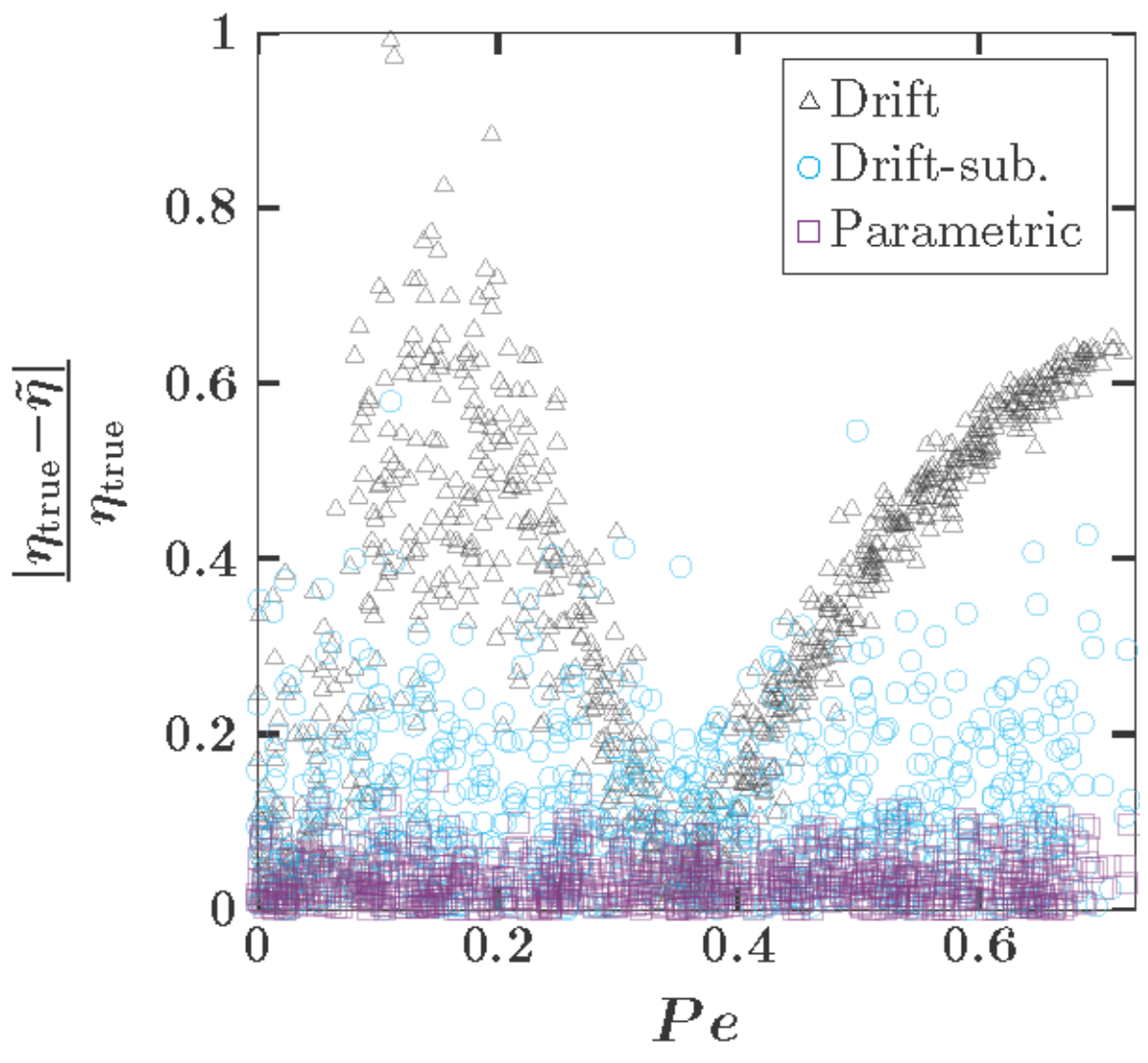

*Figure 5: Relative error in estimates of the viscosity (η) given by the Stokes-Einstein relation based on the three approaches for the Brownian motion data set as a function of the Péclet number (Pe). The mean error in the estimation of η when not accounting for drift is 39.5%. When applying a drift-subtracted least squares approach and a parametric approach the mean error is 11.1% and 3.6%, respectively.*

Figure 6 illustrates the impact of drift on the pathwise estimates of $G'$ and $G''$ for the fBm data for various values of $Pe$. In Figure 7, the ensemble average estimates $G'$ and $G''$ are compared when applying Eqn. (29) to the empirical MSD when ignoring drift and subtracting drift, and applying Eqn. (29) to the parametric scaling of the MSD predicted by our $MLE$ approach for $\Lambda = 9.34 \times 10^{-3}$ μm, corresponding to $Pe = [0.48, 0.52]$ for the fBm data. The ensemble average relative error in the estimation of $G'$ and $G''$ for the fBm data set is reported for the $DLS$ and $MLE$ approaches in Figure 8. *This figure shows that the $MLE$ method more accurately recovers the exact $G'$ and $G''$, uniformly over all frequencies, whereas the $DLS$ method leads to errors in the low frequency range.* The global distortion of individual paths by drift subtraction, the so-called Brownian bridge, foreshadows the modification of path statistics over long lag times, and therefore at low frequencies in transform space.

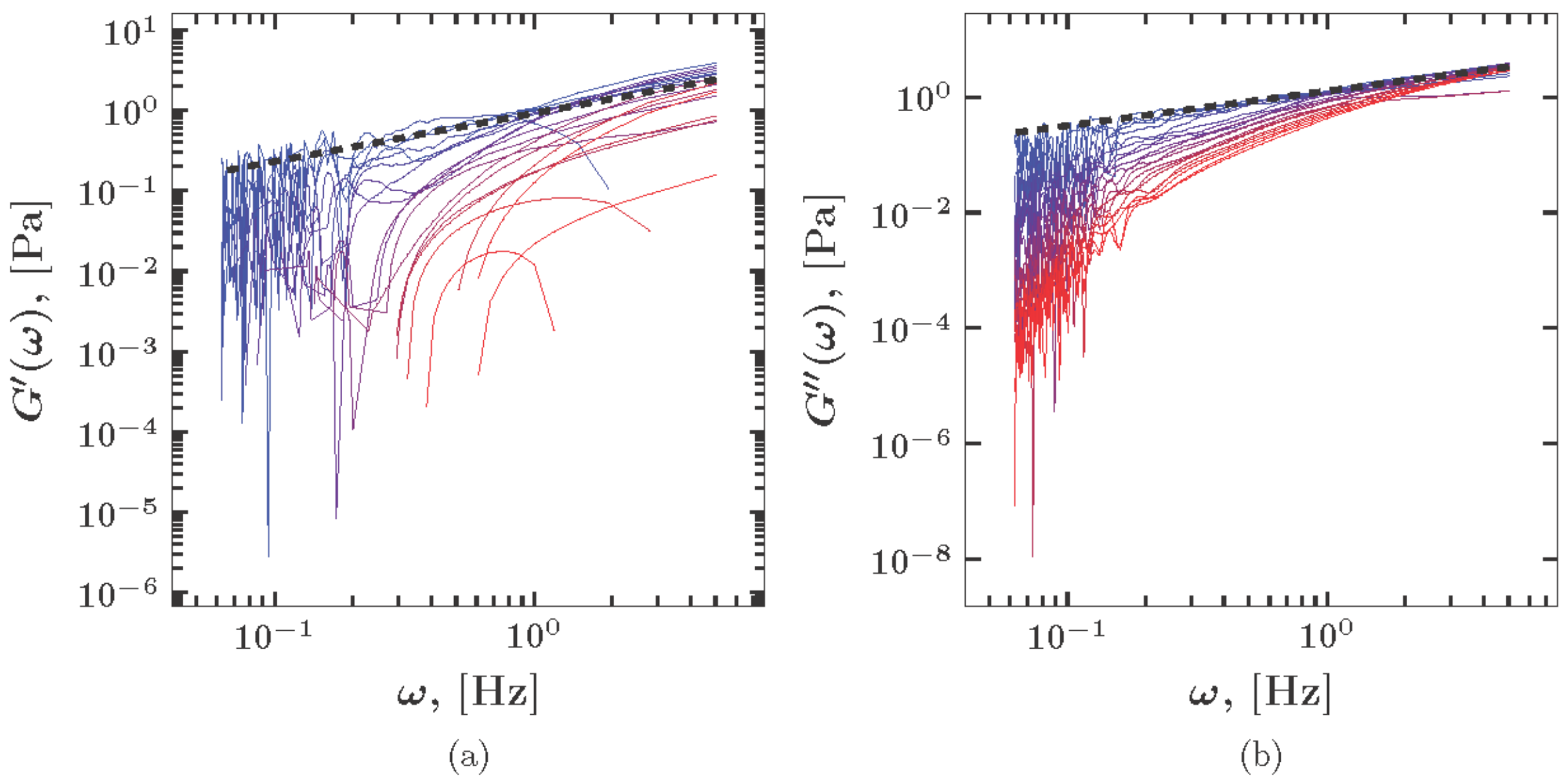

*Figure 6: Pathwise dynamic storage, $G'(\omega)$ (a), and loss, $G''(\omega)$ (b) moduli for the fBm data found by transforming the pathwise MSD without accounting for drift. The change in color of the data corresponds to a transition from $Pe_{min} = 0$ (blue) to $Pe_{max} = 0.52$ (red). The true values of $G'$ and $G''$ are indicated by the black dashed lines.*

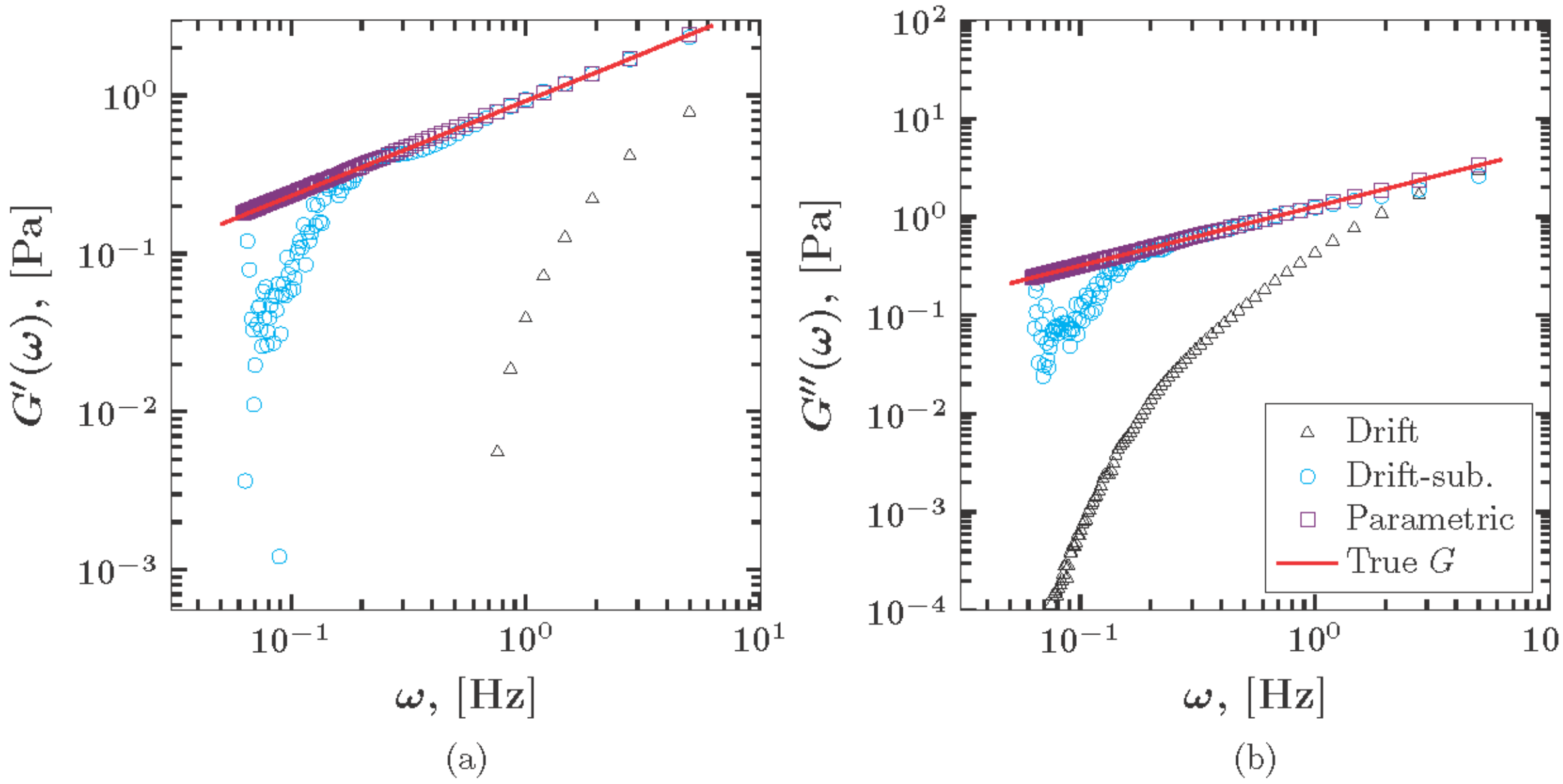

*Figure 7: Ensemble averaged dynamic storage, $G'(\omega)$ (a), and loss, $G''(\omega)$ (b) moduli for the fBm data with subdiffusive exponent $\alpha = 0.6$ by applying Eqn. (29) to the empirical MSD when ignoring drift (black circles) and subtracting drift (blue circles), and applying Eqn. (29) to the parametric scaling of the MSD predicted by our maximum likelihood method (violet circles). Exact $G'(\omega)$ and $G''(\omega)$ are shown in solid red lines. The ensemble-averaged results over 100 paths are shown for $Pe$ values in the range $[0.48, 0.52]$.*

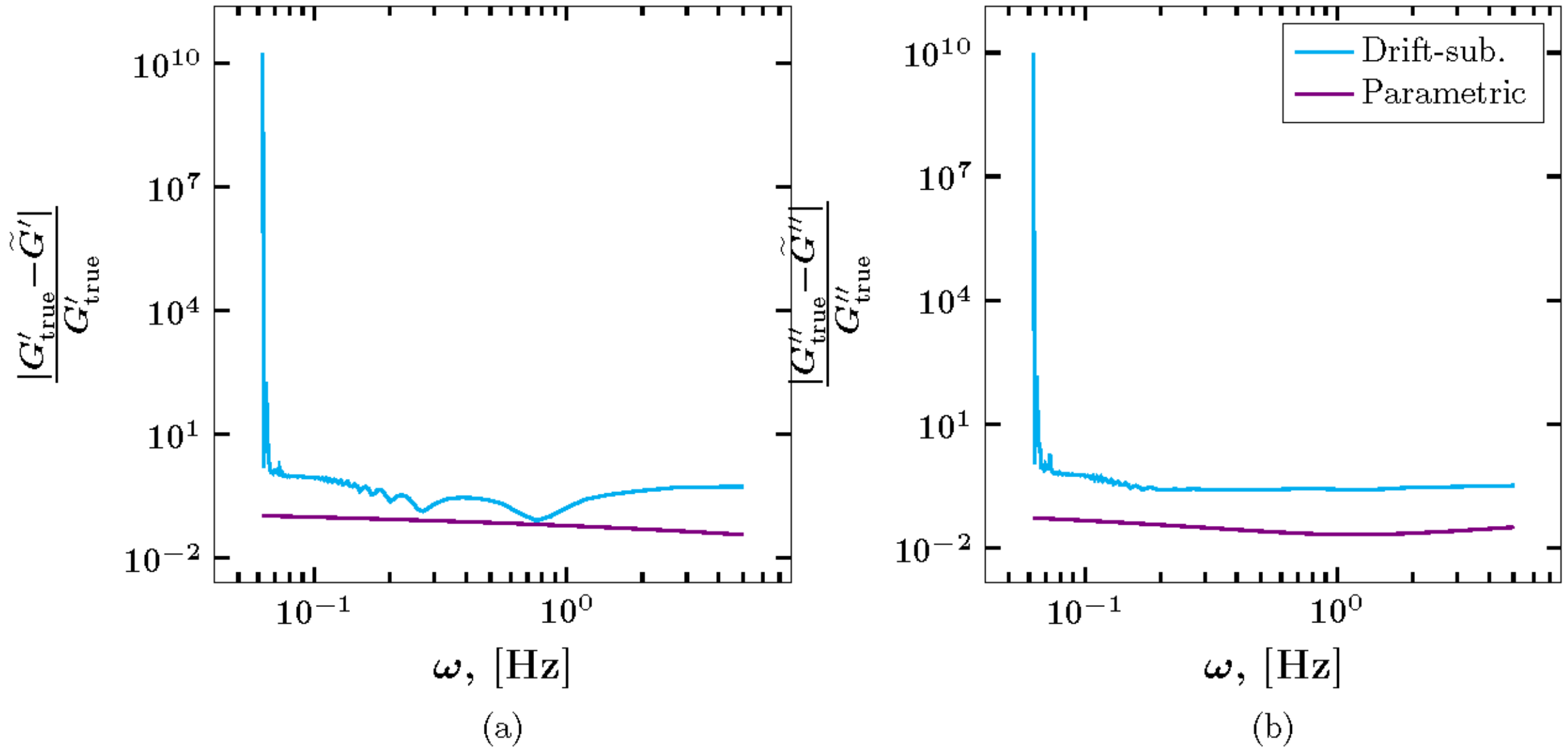

*Figure 8: Ensemble averaged relative error in the storage modulus, $G'(\omega)$ (a), and loss modulus, $G''(\omega)$ (b) for the fBm data with subdiffusive exponent $\alpha = 0.6$ when applying Eqn. (29) to the empirical MSD after subtracting drift (blue) and applying Eqn. (29) to the parametric scaling of the MSD predicted by our maximum likelihood method (violet). The ensemble average is computed using all 4,100 simulated paths.*

We now turn to the underlying challenge to estimate the diffusivity $D$ and the power-law exponent $\alpha$ when the data indicates fractional Brownian motion as a good model. The $LS$, $DLS$, and full model $MLE$ estimates are shown in Figure 9 as a function of the Péclet number. Recall, the true values of each parameter are $\alpha = 0.6$ and $D = 4.67 \times 10^{-4}\ \mu\mathrm{m}^2\mathrm{s}^{-\alpha}$. Failing to account for drift when drift is present leads to highly erroneous results. As $Pe$ increases, $\alpha$ converges to 2, as expected based on the simulation results presented in Figure 1. Drift has a nonlinear impact on the estimate of $D$, initially under-estimating, and later over-estimating the parameter value. In contrast, both the $DLS$ and $MLE$ estimates of $D$ and $\alpha$ are independent of drift and exhibit a similar level of accuracy, however the parametric approach is the more *precise* estimator due to the decreased spread about the mean predicted parameter values.

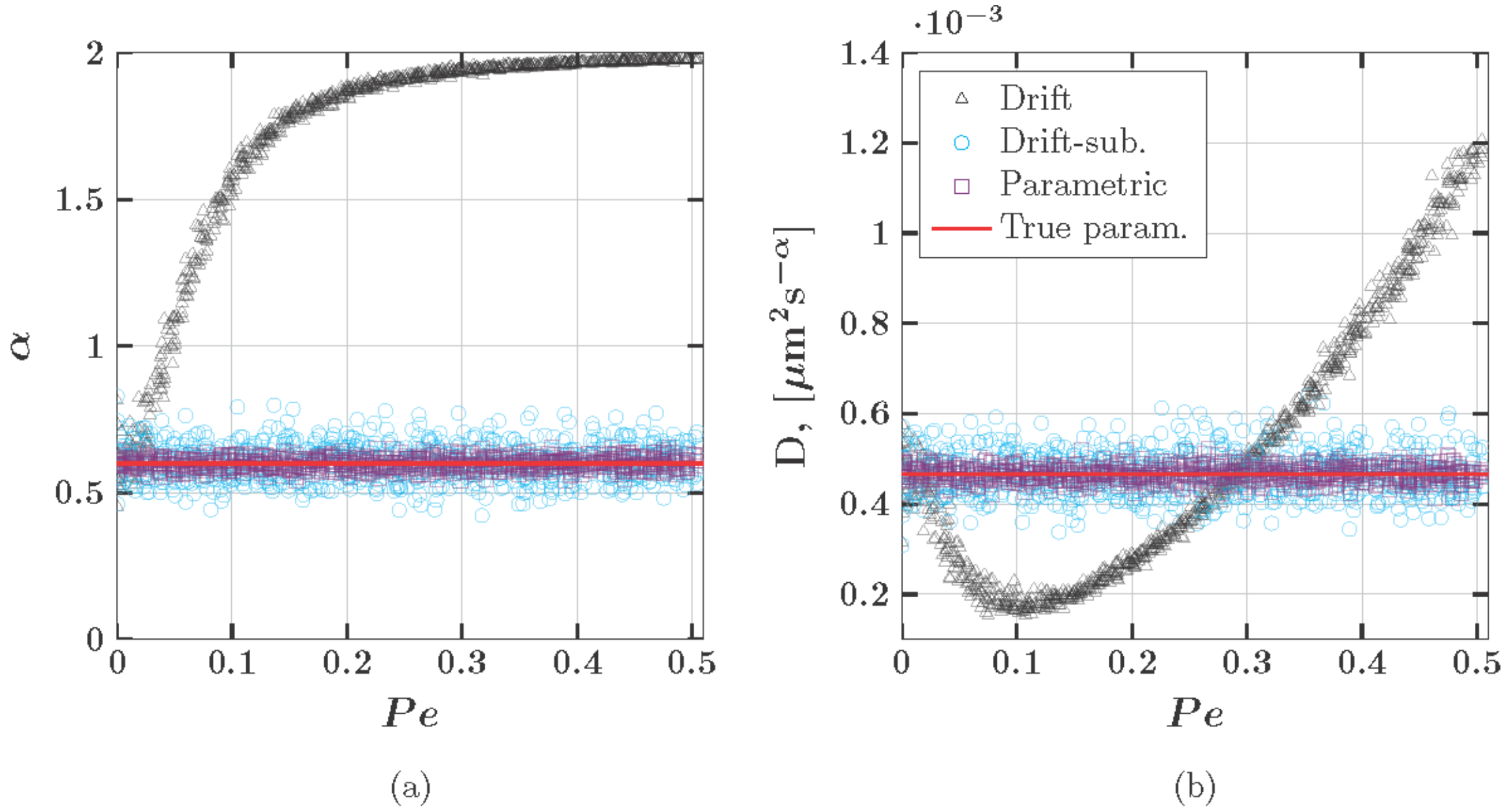

*Figure 9: Estimated values of $\alpha$ (a) and $D$ (b) ignoring drift (black circles), subtracting drift (blue circles) and our maximum likelihood method (violet circles). The true values of $\alpha$ and $D$ are shown in solid red lines. The true values of each parameter are $\alpha = 0.6$ and $D = 4.67 \times 10^{-4}\ \mu m^2 s^{-\alpha}$.*

## VI.B. RESULTS: Experimental Data

Here, we analyze twenty-two representative 1 μm diameter particles in 4 weight percent human bronchial epithelial (HBE) mucus using the previously outlined methods. Each path consists of 2,992 increments with a temporal resolution of 0.2 seconds (five observations per second). The MSD for each experimental path is shown in Figure 10. The ensemble-averaged storage and loss moduli are calculated when ignoring drift, subtracting drift by applying Eqn. (29) to the empirical MSD, and applying Eqn. (29) to the parametric scaling of the MSD of the pure fBm process determined from our maximum likelihood method (Figure 11).

The three estimation methods for $D$ and $\alpha$ were applied to the experimental particle paths. Figure 12 shows the least squares ($LS$) and drift-subtracted least squares ($DLS$) estimates relative to the full model maximum likelihood estimation approach ($MLE$). The estimates for $\alpha$ are presented in Figure 12a. The $MLE$ predictions of $\alpha$ are shown along the x-axis and the $LS$ and $DLS$ estimates are shown on the y-axis. The domain of each axis is from 0, representing stuck particles, to 1, representing normal diffusion exponents. A dashed line indicates the diagonal. Each data point represents the predicted parameter value for one of the 22 experimental paths. A data point falling on the diagonal indicates that the $MLE$ and $LS$ (or $DLS$) approaches agree for that particle path. A data point above the diagonal indicates that the $LS$ (or $DLS$) approach *overestimated* $\alpha$ compared to the $MLE$ approach. Conversely, a data point below the diagonal indicates that the $LS$ (or $DLS$) approach *underestimated* $\alpha$ compared to the $MLE$ approach. The size of each data point is directly proportional to the calculated drift for the corresponding particle.

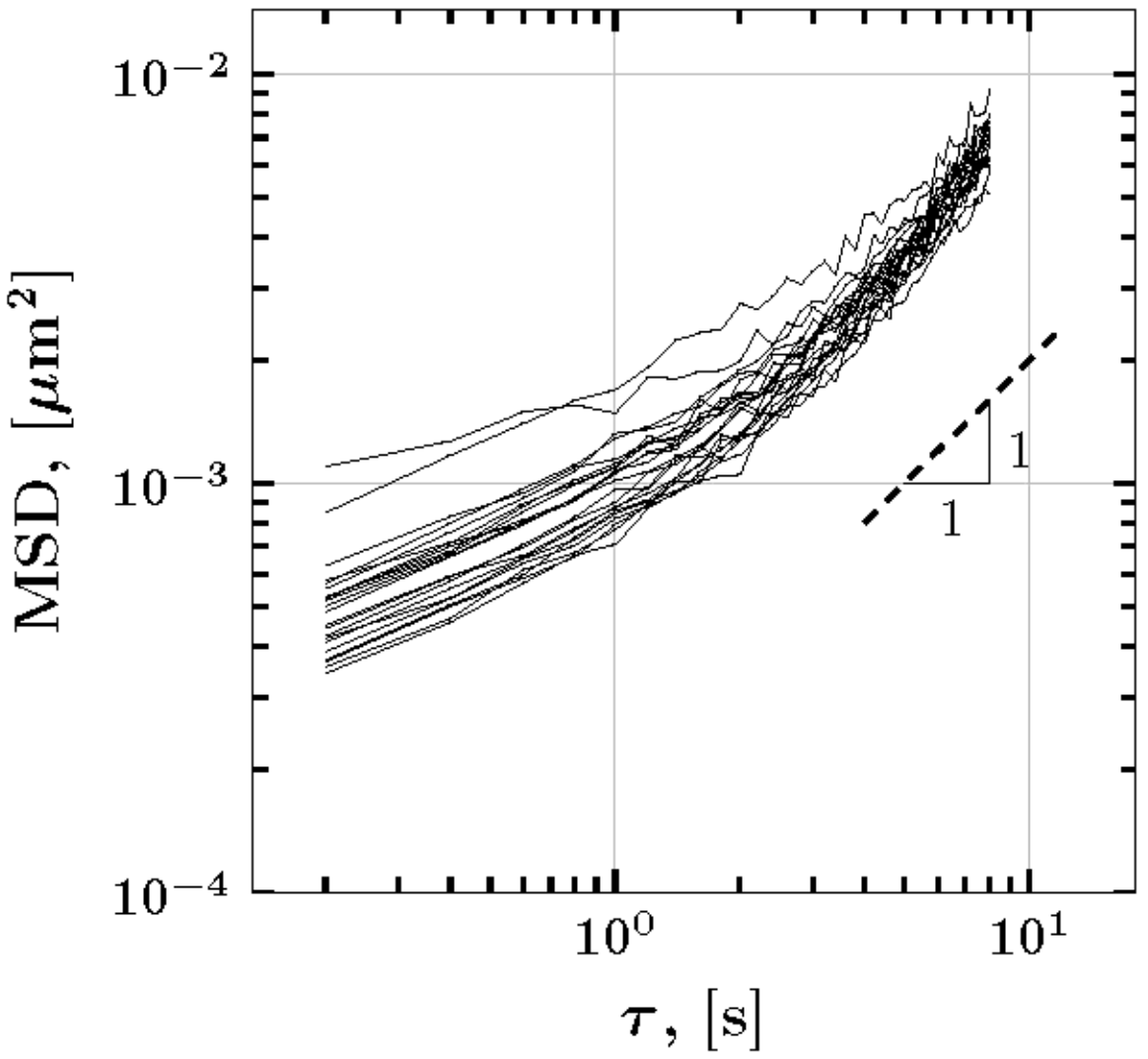

*Figure 10:* Path-wise MSD for the 22 experimental particle paths. The dashed line indicates a slope of 1, corresponding to normal diffusion.

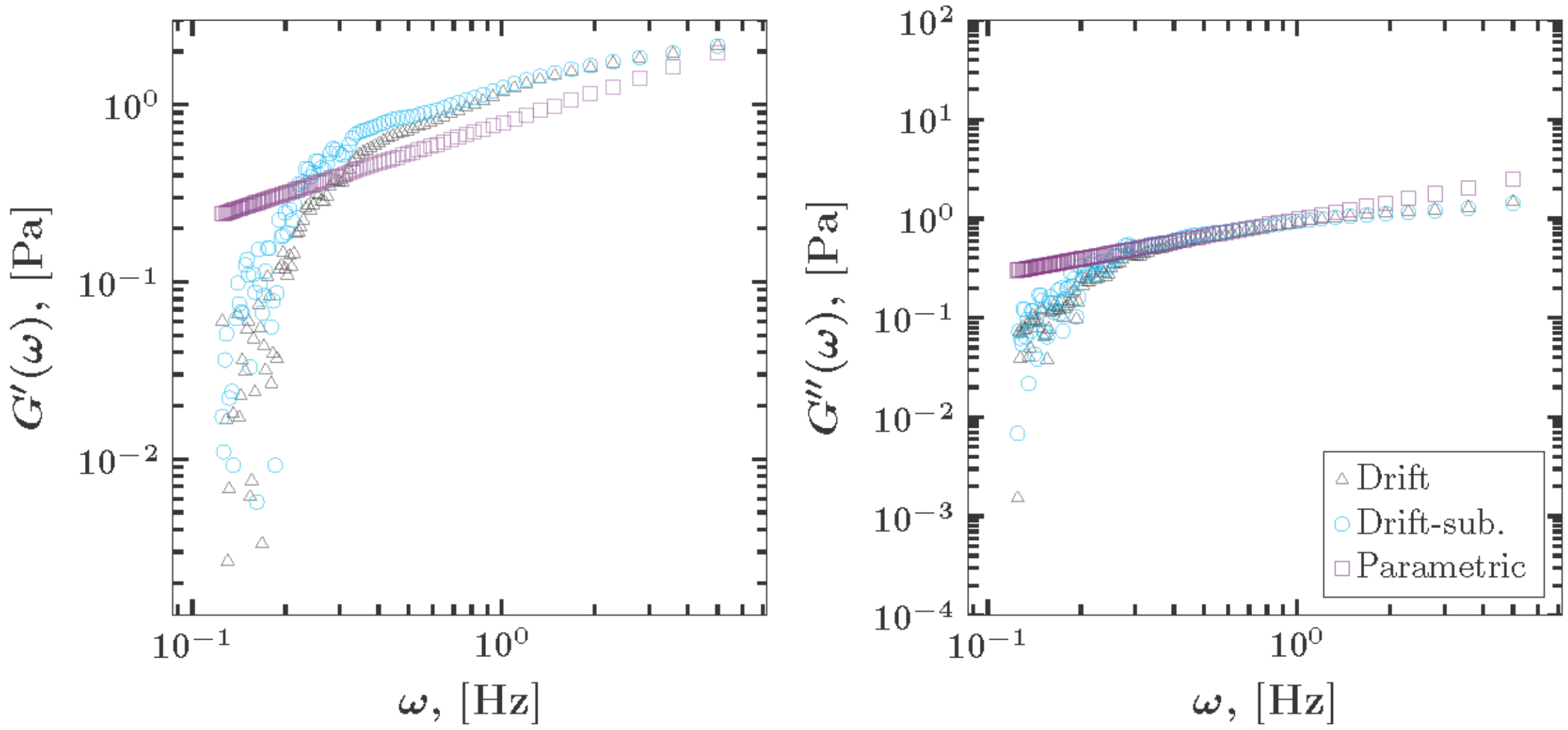

(a) (b)

*Figure 11: Estimates of the dynamic storage modulus, $G'(\omega)$, left figure, and loss modulus, $G''(\omega)$, right figure, for experimental data for 3 approaches: applying Eqn. (29) to the empirical MSD ignoring drift (black circles) and subtracting drift (blue circles), and applying Eqn. (29) to the parametric scaling of the MSD predicted by our maximum likelihood approach (violet circles).*

In Figure 12a, the $DLS$ and $MLE$ approaches exhibit strong agreement in their predictions of $\alpha$, as evidenced by the distribution of data points (blue) along the diagonal dashed line. In contrast, the $LS$ estimates of $\alpha$ (black) fall above the diagonal, indicating an overestimation of $\alpha$ relative to the $MLE$ approach. The amount of overestimation is directly proportional to the amount of drift in the experimental data (larger markers are further from the diagonal than smaller markers).

Figure 12b presents the estimates of the diffusivity ($D$). All data points fall above the diagonal, thus both the $LS$ and $DLS$ approach estimate larger values of $D$ compared with the $MLE$ approach. Here, the amount of overestimation is inversely proportional to the amount of drift (larger markers are closer to the diagonal). We note that this is *not* a scenario observed in the simulated data. Returning to Figure 9, we see that the only time the $LS$ estimates of $D$ are in increasing correspondence to the $MLE$ values with increasing drift is when the $LS$ method underestimates $D$ and $0.1 < Pe < 0.3$. Furthermore, according to Figure 9, the $DLS$ estimate of $D$ relative to the $MLE$ values should be independent of $Pe$. We hypothesize that these incongruences between the simulated and experimental data may be the result of non-linear drift, an interesting issue but beyond the scope of the present paper.

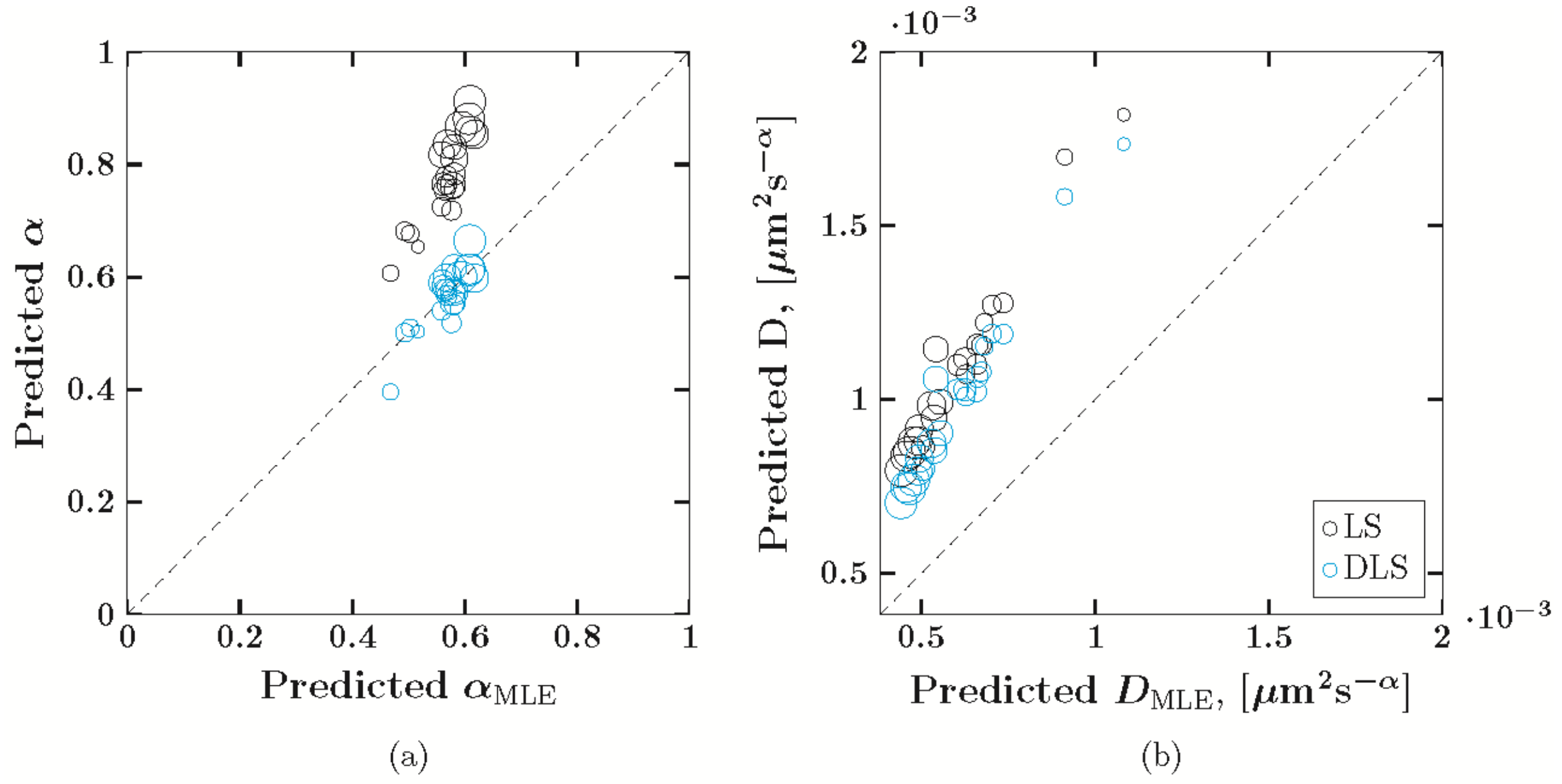

*Figure 12: Ratio of the LS and DLS predictions to the MLE predictions of the power-law exponent $\alpha$ (a) and the prefactor D (b) for experimental data. The LS estimates of both $\alpha$ and D are higher than the MLE estimates. The DLS estimates agree with the MLE estimates for $\alpha$ but are larger for $D$. The size of the marker is proportional to the amount of drift experienced by the particle.*

## VII. DISCUSSION

Persistent linear drift over the course of a particle path is compounded at large lag times, resulting in an asymptotic (long lag time) bending of the MSD curve toward a slope of 2. The use of MSD curves for inference of mobility, creep compliance, or linear viscoelastic moduli, without recognizing there is drift and accounting for it, is obviously problematic. We use a simple calculation of the mean and standard deviation of the step size distribution for a given experimental or numerical particle path to estimate the drift relative to diffusion, defining an approximate Péclet number $Pe$.

When $Pe$ increases, the slope of the path-wise MSD approaches 2 at increasingly smaller lag times, causing the least squares ($LS$) estimate of the power law exponent, $\alpha$, to converge to 2. By subtracting the mean increment of each particle from the particle's path, the least squares ($DLS$) estimate of each parameter is more stable. However, the unanticipated correlation in the increment process induced by drift subtraction over increasingly large lag times leads to an error in the estimation of the diffusive parameters that is on the order of 10%. Accordingly, these discrepancies at large lag times produce errors in creep compliance at those lag times and in the dynamic moduli at low frequencies.

To address this issue, we advocate for a parametric maximum likelihood estimation ($MLE$) approach that identifies the best-fit drift parameter $\mu$ simultaneously while estimating $D$ and $\alpha$. We demonstrate on numerically generated particle paths, with physically relevant diffusive parameters from human bronchial epithelial mucus studies, that the use of the parametric maximum likelihood approach results in approximately a 2/3 reduction in the error in the exact fractional Brownian motion parameters $D$ and $\alpha$ compared to the standard drift-subtraction least squares approach. Such diffusive mobility parameters are routinely used in drug delivery to compare various drug delivery particle formulations for passage through mucosal layers [7, 8, 10]. With respect to inference of linear viscoelasticity from the MSD statistics of particle paths, we have illustrated that accuracy in storage and loss moduli deteriorates at low frequencies for the standard drift-subtraction, least squares methods. The gains in accuracy by the $MLE$ method have been shown for fractional Brownian numerical data typical of experimentally observed data in mucus gels. Furthermore, we note that the statistical properties of the $MLE$ method are well understood in the statistics community, and have further value beyond that illustrated here, e.g., for testing model assumptions against experimental data as in [21].

To close, the $MLE$, least squares and drift-subtracted least squares parameter estimation approaches were applied to experimental paths of 1 μm diameter beads in human bronchial epithelial cell culture mucus [9]. Relative to the parametric maximum likelihood approach, the drift-subtracted least squares method predicts a higher elasticity above $\sim 0.3$ Hz (thus biasing toward more gel-like properties at these frequencies [53]), and a lower viscosity for all frequencies. The comparison of the $MLE$ approach with macrorheology data for inference of the creep compliance or dynamic moduli of HBE culture mucus is a current project of interest. The sensitivity of HBE mucus to weight percent solids [9] and the extremely low stress thresholds for linear response (unpublished data) make this macro-micro rheology comparison more intricate than for other systems such as actin filament networks [1].

## VII. ACKNOWLEDGEMENTS

The authors gratefully acknowledge partial support from the National Science Foundation Grants DMS-1412844, DMS-1100281, DMS-1462992, DMS-1412998, DMS-1410047, DMS-1107070, National Institutes of Health and National Heart Lung and Blood Institute grants NIH/NHLBI 1 P01 HL108808-01A1, NIH/NHLBI 5 R01 HL 077546-05, and Natural Sciences and Engineering Research Council of Canada grant RGPIN-2014-04225.

## Appendix. Pseudocode for MLE of fBM with Drift

*Instructions:*

1. For a given set of observations $\boldsymbol{X} = (X_0, \dots, X_M)$ with $X_i = X(i\Delta t)$, calculate the increments

$\boldsymbol{x} = (x_1, \dots x_M)$, $x_i = x_i - x_{i-1}$.

2. For given $\boldsymbol{x}$ and $\Delta t$, maximize the function $\textsc{ProfLogLik}(\alpha, \boldsymbol{x}, \Delta t)$ as a function of α. For fBM, we have $0 < \alpha < 2$, so the maximum can easily be obtained using the fminbnd routine in MATLAB.

3. Once we have the *MLE* $\hat{\alpha}$ of $\alpha$, we can find the *MLE*s of $\mu$ and $\sigma^2$ with $\{\hat{\mu}, \hat{\sigma}^2\} = \textsc{mle}(\hat{\alpha}, \boldsymbol{x}, \Delta t)$.

➢ Evaluate $\ell_{\text{prof}}(\alpha|\boldsymbol{x})$

**function** $\textsc{ProfLogLik}(\alpha, \boldsymbol{x}, \Delta t)$

$M \leftarrow \textbf{length}(\boldsymbol{x})$

$\{\hat{\mu}_\alpha, \hat{\sigma}_\alpha^2, \nu\} \leftarrow \textsc{mle}(\alpha, \boldsymbol{x}, \Delta t)$

$\lambda \leftarrow -\frac{1}{2}(M\ln(\hat{\sigma}_\alpha^2) + \nu)$

**return** $\lambda$

**end function**

- Calculate the *MLE*s $\hat{\mu}_\alpha$ and $\hat{\sigma}_\alpha^2$, as well as $\nu = \ln(|\boldsymbol{V}_\alpha|)$

**function** $\text{MLE}(\alpha, \boldsymbol{x}, \Delta t)$

$\boldsymbol{\gamma}_{M\times 1} \leftarrow \text{ACF}(\alpha, \Delta t, M)$

$\boldsymbol{A}_{2\times M} \leftarrow \begin{bmatrix} x_1 & \cdots & x_M \\ \Delta t & \cdots & \Delta t \end{bmatrix}$

$\{\boldsymbol{Q}_{2\times 2}, \nu\} \leftarrow \text{DurbinLevinson}(\boldsymbol{\gamma}, \boldsymbol{A})$

$\hat{\mu}_\alpha \leftarrow Q_{12}/Q_{22}$

$\hat{\sigma}_\alpha^2 \leftarrow \frac{1}{M}(Q_{11} - Q_{12}\hat{\mu}_\alpha)$

**return** $\{\hat{\mu}_\alpha, \hat{\sigma}_\alpha^2, \nu\}$

**end function**

- Autocorrelation of fBm increments

**function** $\text{ACF}(\alpha, \Delta t, M)$

$\boldsymbol{\gamma}_{M\times 1} \leftarrow \boldsymbol{0}_{M\times 1}$

**for** $i \leftarrow 1{:}M$ **do**

$\gamma_i \leftarrow \frac{1}{2}\Delta t^\alpha(|i+1|^\alpha + |i-1|^\alpha - 2|i|^\alpha)$

**end for**

**return** $\boldsymbol{\gamma}$

**end function**

➢ Calculate $\boldsymbol{Q} = \boldsymbol{A}\boldsymbol{V}^{-1}\boldsymbol{A}'$ and $\nu = \ln(|\boldsymbol{V}|)$, where $\boldsymbol{V} = \textsc{Toeplitz}(\boldsymbol{\gamma})$

```
function DurbinLevinson(γ, A)
    {d, M} ← dims(A_{d×M})
    ν ← 0
    Q_{d×d} ← 0_{d×d}
    φ_{M×1} ← 0_{M×1}
    τ ← γ_1
    for i ← 0:(M − 1) do
        z_{d×1} ← A_{(1:d,i+1)}                      % Indices subset blocks of matrices
        if i > 0 then
            z ← z − A_{(1:d,1:i)} φ_{i:1}            % Reverse elements in φ_{1:i}
            τ ← τ(1 − φ_i^2)
        end if
        Q ← Q + zz'/τ                                % Outer product of z with itself
        ν ← ν + log(τ)
        if i < M − 1 then
            φ_{i+1} ← (γ_{i+2} − φ'_{1:i} γ_{i+1:2})/τ   % Inner product of φ_{1:i} and γ_{i+1:2}
            φ_{1:i} ← φ_{1:i} − φ_{i+1} φ_{i:1}
        end if
    end for
    return {Q, ν}
end function
```

## REFERENCES

[1] Xu, J., V. Viasnoff and D. Wirtz, "Compliance of actin filament networks measured by particle-tracking microrheology and diffusing wave spectroscopy", Rheol. Acta **37**, 387-398 (1998).

[2] Mason, T.G. and D.A. Weitz, "Optical measurements of frequency-dependent linear viscoelastic moduli of complex fluids", Phys. Rev. Lett. **74**, 1250-1253 (1995).

[3] Mason, T.G. and D.A. Weitz, "Linear viscoelasticity of colloidal hard sphere suspensions near the glass transition", Phys. Rev. Lett. **75**, 2770-2773 (1995).

[4] Mason, T.G., "Estimating the viscoelastic moduli of complex fuids using the generalized Stokes-Einstein equation", Rheol. Acta **39**, 371-378 (2000).

[5] Dasgupta, B.R., S.Y. Tee, J.C. Crocker, B.J. Frisken and D.A. Weitz, "Microrheology of polyethylene oxide using diffusing wave spectroscopy and single scattering", Phys. Rev. E Stat. Nonlin. Soft Matter Phys. **65**, 51505 (2002).

[6] das Neves, J., C.M.R. Rocha, M.P. Gonçalves, R.L. Carrier, M. Amiji, M.F. Bahia and B. Sarmento, "Interactions of microbicide nanoparticles with a simulated vaginal fluid", Mol. Pharm. **9**, 3347-56 (2012).

[7] Lai, S.K., Y.-Y. Wang, R. Cone, D. Wirtz and J. Hanes, "Altering mucus rheology to "solidify" human mucus at the nanoscale", PLOS ONE **4**, e4294 (2009).

[8] Wang, Y.-y., S.K. Lai, L.M. Ensign, W. Zhong, R. Cone and J. Hanes, "The microstructure and bulk rheology of human cervicovaginal mucus are remarkably resistant to changes in pH", Biomacromolecules **14**, 4429-35 (2013).

[9] Hill, D.B., P.A. Vasquez, J. Mellnik, S.A. McKinley, A. Vose, F. Mu, A.G. Henderson, S.H. Donaldson, N.E. Alexis, R.C. Boucher and M.G. Forest, "A biophysical basis for mucus solids concentration as a candidate biomarker for airways disease", PLOS ONE **9**, e87681 (2014).

[10] Schuster, B.S., L.M. Ensign, D.B. Allan, J.S. Suk and J. Hanes, "Particle tracking in drug and gene delivery research: state-of-the-art applications and methods", Adv. Drug Deliver. Rev. **91**, 70-91 (2015).

[11] Georgiades, P., P.D.A. Pudney, S. Rogers, D.J. Thornton and T.A. Waigh, "Tea derived galloylated polyphenols cross-link purified gastrointestinal mucins", PLOS ONE **9**, e105302 (2014).

[12] Georgiades, P., P.D.A. Pudney, D.J. Thornton and T.A. Waigh, "Particle tracking microrheology of purified gastrointestinal mucins", Biopolymers **101**, 366-77 (2014).

[13] Macierzanka, A., A.R. Mackie, B.H. Bajka, N.M. Rigby, F. Nau and D. Dupont, "Transport of particles in intestinal mucus under simulated infant and adult physiological conditions: impact of mucus structure and extracellular DNA", PLOS ONE **9**, e95274 (2014).

[14] Adler, J. and S.N. Pagakis, "Reducing image distortions due to temperature-related microscope stage drift", J. Microsc. **210**, 131-137 (2003).

[15] Dangaria, J.H., S. Yang and P.J. Butler, "Improved nanometer-scale particle tracking in optical microscopy using microfabricated fiduciary posts", Biotechniques **42**, 437-440 (2007).

[16] Aufderhorst-Roberts, A., W.J. Frith, M. Kirkland and A.M. Donald, "Microrheology and microstructure of Fmoc-derivative hydrogels", Langmuir **30**, 4483-4492 (2014).

[17] Savin, T. and P.S. Doyle, "Static and dynamic errors in particle tracking microrheology", Biophys. J. **88**, 623-638 (2005).

[18] Hasnain, I.A. and A.M. Donald, "Microrheological characterization of anisotropic materials", Phys. Rev. E Stat. Nonlin. Soft Matter Phys. **73**, 31901 (2006).

[19] Fong, E.J., Y. Sharma, B. Fallica, D.B. Tierney, S.M. Fortune and M.H. Zaman, "Decoupling directed and passive motion in dynamic systems: particle tracking microrheology of sputum", Ann. Biomed. Eng. **41**, 837-846 (2013).

[20] Metzler, R. and J. Klafter, "The restaurant at the end of the random walk: recent developments in the description of anomalous transport by fractional dynamics", J. Phys. A Math. Gen. **37**, R161--R208 (2004).

[21] Lysy, M., N.S. Pillai, D.B. Hill, M.G. Forest, J. Mellnik, P. Vazquez and S.A. McKinley, "Model comparison and assessment for single particle tracking in biological fluids", J. Am. Stat. Assoc. (accepted) (2016).

[22] Einstein, A., "On the movement of small particles suspended in stationary liquids required by the molecular-kinetic theory of heat", Ann. Phys. Leipzig **17**, 549-560 (1905).

[23] Qian, H., M.P. Sheetz and E.L. Elson, "Single particle tracking. Analysis of diffusion and flow in two-dimensional systems", Biophys. J. **60**, 910-921 (1991).

[24] Michalet, X., "Mean square displacement analysis of single-particle trajectories with localization error: Brownian motion in an isotropic medium", Phys. Rev. E Stat. Nonlin. Soft Matter Phys. **82**, 41914 (2010).

[25] Gal, N., D. Lechtman-Goldstein and D. Weihs, "Particle tracking in living cells: a review of the mean square displacement method and beyond", Rheol. Acta **52**, 425-443 (2013).

[26] Weihs, D., M.A. Teitell and T.G. Mason, "Simulations of complex particle transport in heterogeneous active liquids", Microfluid. Nanofluid. **3**, 227-237 (2007).

[27] Kou, S.C. and X.S. Xie, "Generalized Langevin Equation with fractional Gaussian noise: subdiffusion within a single protein molecule", Phys. Rev. Lett. **93**, 1-4 (2004).

[28] Kou, S.C., "Stochastic modeling in nanoscale biophysics: subdiffusion within proteins", Ann. Appl. Stat. **2**, 501-535 (2008).

[29] Caspi, A., R. Granek and M. Elbaum, "Enhanced diffusion in active intracellular transport", Phys. Rev. Lett. **85**, 5655-5658 (2000).

[30] Seisenberger, G., M.U. Ried, T. Endreß, H. Büning, M. Hallek and C. Bräuchle, "Real-time single-molecule imaging of the infection pathway of an Adeno-associated virus", Science **294**, 1929-1933 (2001).

[31] Valentine, M.T., P.D. Kaplan, D. Thota, J.C. Crocker, T. Gisler, R.K. Prud'homme, M. Beck and D.A. Weitz, "Investigating the microenvironments of inhomogeneous soft materials with multiple particle tracking", Phys. Rev. E Stat. Nonlin. Soft Matter Phys. **64**, 61506 (2001).

[32] Steele, J.M., *Stochastic calculus and financial applications*, (Springer, New York, 2001) .

[33] Mandelbrot, B.B. and J.W. Van Ness, "Fractional Brownian motions, fractional noises and applications", SIAM Rev. **10**, 422-437 (1968).

[34] Fulcher, M.L., S. Gabriel, K.A. Burns, J.R. Yankaskas and S.H. Randell, "Well-differentiated human airway epithelial cell cultures", Methods Mol. Med. **107**, 183-206 (2005).

[35] Matsui, H., M.W. Verghese, M. Kesimer, U.E. Schwab, S.H. Randell, J.K. Sheehan, B.R. Grubb and R.C. Boucher, "Reduced three-dimensional motility in dehydrated airway mucus prevents neutrophil capture and killing bacteria on airway epithelial surfaces", J. Immun. **175**, 1090-1099 (2005).

[36] Matsui, H., V.E. Wagner, D.B. Hill, U.E. Schwab, T.D. Rogers, B. Button, R.M. Taylor, R. Superfine, M. Rubinstein, B.H. Iglewski and R.C. Boucher, "A physical linkage between cystic fibrosis airway surface dehydration and Pseudomonas aeruginosa biofilms", Proc. Natl. Acad. Sci. USA **103**, 18131-6 (2006).

[37] Hill, D.B. and B. Button, "Establishment of respiratory air-liquid interface cultures and their use in studying mucin production, secretion, and function", in *Mucins: Methods and Protocols*, edited by M.A. McGucking and D.J. Thornton (Springer, New York, 2012) , pp. 245-258.

[38] Kesimer, M., S. Kirkham, R.J. Pickles, A.G. Henderson, N.E. Alexis, G. Demaria, D. Knight, D.J. Thornton and J.K. Sheehan, "Tracheobronchial air-liquid interface cell culture: a model for innate mucosal defense of the upper airways?", Am. J. Physiol. Lung Cell. Mol. Physiol. **296**, L92--L100 (2009).

[39] Button, B., L.-H. Cai, C. Ehre, M. Kesimer, D.B. Hill, J.K. Sheehan, R.C. Boucher and M. Rubinstein, "A periciliary brush promotes the lung health by separating the mucus layer from airway epithelia", Science **337**, 937-41 (2012).

[40] Anderson, W.H., R.D. Coakley, B. Button, A.G. Henderson, L. Kirby, N.E. Alexis, D.B. Peden, E.R. Lazarowski, C.W. Davis, S. Bailey, F. Fuller, M. Almond, B. Qaqish and E. Bordonali, "The

relationship of mucus concentration (hydration) to mucus osmotic pressure and transport in chronic bronchitis", Am. J. Resp. Crit. Care **192**, 182-190 (2015).

[41] Lai, S.K., D.E. O'Hanlon, S. Harrold, S.T. Man, Y.-Y. Wang, R. Cone and J. Hanes, "Rapid transport of large polymeric nanoparticles in fresh undiluted human mucus", Proc. Natl. Acad. Sci. USA **104**, 1482-7 (2007).

[42] Cone, R.A., "Barrier properties of mucus", Adv. Drug. Deliver. Rev. **61**, 75-85 (2009).

[43] Lai, S.K., Y.-Y. Wang, D. Wirtz and J. Hanes, "Micro-and macrorheology of mucus", Adv. Drug. Deliver. Rev. **61**, 86-100 (2009).

[44] Lai, S.K., Y.-Y. Wang, K. Hida, R. Cone and J. Hanes, "Nanoparticles reveal that human cervicovaginal mucus is riddled with pores larger than viruses", Proc. Natl. Acad. Sci. USA **107**, 598-603 (2010).

[45] Dawson, M., D. Wirtz and J. Hanes, "Enhanced viscoelasticity of human cystic fibrotic sputum correlates with increasing microheterogeneity in particle transport", J. Bio. Chem. **278**, 50393-401 (2003).

[46] Schuster, B.S., J.S. Suk, G.F. Woodworth and J. Hanes, "Nanoparticle diffusion in respiratory mucus from humans without lung disease", Biomaterials **34**, 3439-3446 (2013).

[47] Waigh, T.A., "Microrheology of complex fluids", Rep. Progr. Phys. **68**, 685-742 (2005).

[48] Levy, R., D.B. Hill, M.G. Forest and J.B. Grotberg, "Pulmonary fluid flow challenges for experimental and mathematical modeling", Integr. Comp. Biol. **54**, 985-1000 (2014).

[49] Davidson, A.C., *Statistical models*, (Cambridge University Press, New York, 2003) .

[50] Bareiss, E.H., "Numerical solution of linear equations with Toeplitz and vector Toeplitz matrices", Numer. Math. **13**, 404-424 (1969).

[51] Durbin, J., "The fitting of time-series models", Rev. Inst. Int. Stat. **28**, 233-244 (1960).

[52] Ljung, L., *System identification: theory for the user*, (Prentice-Hall, Englewood Cliffs, 1987) .

[53] Winter, H.H., "Can the gel point of a crosslinking polymer be detected by the G'-G" crossover?", Polym. Eng. Sci. **27**, 1698-1702 (1987).